\documentclass[aps%
,prb%
,twocolumn%
,showpacs%
,floats
,amssymb%
]{revtex4}
\usepackage{amsmath}%
\usepackage{graphicx}%
\usepackage{color}%

\newcommand{\comment}[1]{}

\begin{document}
\title{Reduction of Dilute Ising Spin Glasses} 
\date{\today}
\author{Stefan Boettcher}
\homepage{http://www.physics.emory.edu/faculty/boettcher/}
\author{James Davidheiser}  
\homepage{http://www.physics.emory.edu/students/davidheiser/}
\affiliation{Physics Department, Emory University, Atlanta, Georgia
30322, USA}    

\begin{abstract} 
The recently proposed reduction method for diluted spin glasses is
investigated in depth. In particular, the Edwards-Anderson model with
$\pm J$ and Gaussian bond disorder on hyper-cubic lattices in $d=2$,
3, and 4 is studied for a range of bond dilutions. The results
demonstrate the effectiveness of using bond dilution to elucidate
low-temperature properties of Ising spin glasses, and provide a
starting point to enhance the methods used in reduction. Based on
that, a new greedy heuristic call ``Dominant Bond Reduction'' is
introduced and explored.
\end{abstract} 

\pacs{%
02.60.Pn
, 05.50.+q
, 75.10.Nr
.}
\maketitle

\section{Introduction}
\label{intro}
Despite more than three decades of intensive research, many properties
of spin glasses~\cite{Binder86,MPV,F+H,Young98}, especially
in finite dimensions, are still not well understood. The most simple
model is the Edwards-Anderson model (EA)~\cite{Edwards75},
\begin{eqnarray}
H=-\sum_{<i,j>}\,J_{i,j}\,x_i\,x_j,\quad(x_i=\pm1),
\label{Heq}
\end{eqnarray}
with Ising spins $x_i=\pm1$ arranged on a finite-dimensional lattice
with nearest-neighbor bonds $J_{i,j}$, randomly drawn from a
distribution $P(J)$ of zero mean and unit variance.

In Refs.~\onlinecite{Boettcher04c,Boettcher04b,MKpaper}, it was proposed to
study the EA in Eq.~(\ref{Heq}) on bond-diluted lattices at $T=0$ to
obtain more accurate scaling behavior for low-temperature
excitations. There, it is shown how to remove iteratively
low-connected spins from the lattice and alter the interactions,
i.~e., to {\em reduce} the system, in such a way that the ground-state
energy of the reduced system is the same as the original system. In
this way often much larger lattice sizes $L$ can be simulated compared
to undiluted ones and, in combination with finite-size scaling,
enhanced scaling regimes are achieved. With these methods, for
instance, we have solved spin glasses exactly at the bond-percolation
threshold $p_c$, the edge of the glassy regime, in $d=2,\ldots,7$ by
reducing a large number of systems with up to $10^8$
spins~\cite{Boettcher07a}. 

There is, of course, a long history of studying spin systems
  on dilute lattices, including spin glasses, going back to
  Ref.~\onlinecite{Edwards75} itself; see for example
  Refs.~\onlinecite{Stephen77,Southern79,Kolan82,Banavar87,Bray87b}.
  Coniglio and co-workers have proposed to map the ensemble of
  critical Ising (or Potts) spin models~\cite{Coniglio89,Coniglio99a}
  onto percolating clusters, based on the ideas of Fortuin and
  Kasteleyn~\cite{Fortuin72}, to treat ferromagnetic~\cite{Coniglio96}
  and spin glass phenomena~\cite{Coniglio99b,Coniglio99c}. Our
  approach here is based on transformations in the Hamiltonian of an
  Ising spin system that are \emph{exact} for each instance. The price
  paid is that these transformations  reducing the Hamiltonian only
  apply at $T=0$. Extending our earlier work on the
  Migdal-Kadanoff approximation~\cite{MKpaper}, Ref.~\onlinecite{Jorg08}
  very recently included infinitesimal temperature corrections. Their
  method is limited to discrete bonds and a subset of the rules we
  discuss here. As we can merely consider $T=0$, we are unfortunately
  not sensitive to the novel transition seen by Ref.~\onlinecite{Jorg08}
  that is said to emerge only at non-zero temperatures.  

Our approach, combined with the highly efficient ``Extremal
Optimization'' (EO)
heuristic~\cite{Boettcher01a,Boettcher00}, have lead to a
comprehensive characterization of low-temperature excitations in spin
glasses in low dimensions (up to $d=7$)~\cite{Boettcher05d}. These
results allow for a direct comparison with mean-field
predictions~\cite{Aspelmeier03}, and have recently motivated a
re-assessment of fundamental scaling
relations~\cite{Parisi07,Aspelmeier07}. This work has also inspired
the use of dilution for more effective Monte Carlo simulations of
disordered systems~\cite{Jorg04,Jorg06,Jorg08,Hasenbusch08}.

Here we study the connection between lattice-topology and the
reduction method. In particular, we explore the effectiveness of each
of the reduction rules as a function of bond dilution. The results
provides the reader with an opportunity to evaluate in more detail the
conclusions drawn in previous
papers~\cite{Boettcher03a,Boettcher03b,Boettcher04c,Boettcher04b,Boettcher05d},
and might suggest possible extensions of these rules for improved
effectiveness. As an example of a concrete application, we introduce
and evaluate ``Dominant Bond Reduction'' (DBR), a new heuristic which
provides a greedy, almost linear algorithm to obtain approximate spin-glass ground
states on average with bounded relative error for increasing system sizes.

This paper is structured as follows. In the next Section, we will
introduce the reduction method and its rules. In Sec.~\ref{numerics},
we display and discuss our numerical investigation of the
reduction rules. In Sec.~\ref{DBR}, we discuss our simulation results
for DBR, followed by some concluding remarks in
Sec.~\ref{conclusions}. In the Appendix, a generalized description of
the reduction method is provided, with some speculations on its
applicability.

\section{Reduction Method}
\label{reduction}
To exploit the advantages of spin glasses on a bond-diluted lattice,
we can often {\it reduce} the number of relevant degrees of freedom
substantially before a call to an optimization algorithm becomes
necessary. Such a reduction, in particular of low-connected spins,
leads to a smaller, compact remainder graph, bare of trivially
fluctuating variables, which is easier to optimize. These reductions
are very similar to a procedure known as k-core decomposition in graph
theory, which is often applied to analyze hard combinatorial or
real-world
problems~\cite{Farrow05,Vespignani06,Farrow07}. Furthermore, rules of
this sort have also been used to study planar~\cite{Frank88} and $3d$
resistor networks~\cite{Gingold90}.

Here, we focus exclusively on the reduction rules for the ground-state
energy (i.~e., $T=0$); a subset of these rules also permit the exact
determination of the entropy and overlap~\cite{MKpaper} at
$T=0$. These rules apply to general Ising spin glass Hamiltonians as
in Eq.~(\ref{Heq}) with {\it any} bond distribution $P(J)$, discrete
or continuous, on arbitrary sparse graphs.

The reductions affect both spins and bonds, eliminating recursively
all zero-, one-, two-, and three-connected spins. These rules are
supplemented here with one that is not topological but concerns bond
values directly, which is especially effective for broad, continuous
bond distributions, like Gaussian or power-law bonds. The addition of
more elaborate rules that depend of suitably chosen bond distributions
is conceivable, the universality of the underlying physics permitting.
These operations eliminate and add terms to the expression for the
Hamiltonian in Eq.~(\ref{Heq}), but leave it \emph{form-invariant.}
Again, loosening this requirement may lead to an even more efficient
procedure for certain problems, although it should be understood that
an unrestricted reduction in general leads to an exponential growth in
the number and form of the newly created terms, involving all
combinations of $p$-spin interactions (see the Appendix).  Offsets in
the energy along the way are accounted for by a variable $H_o$, which
is {\it exact} for a ground-state configuration. The rules discussed
here are as follows:

{\it Rule~I:} An isolated spin can be ignored entirely.

{\it Rule~II:} A one-connected spin $i$ can be eliminated, since its
state can always be chosen in accordance with its neighboring spin $j$
to satisfy the bond $J_{i,j}$. For its energetically most favorable
state we adjust $H_o:=H_o-|J_{i,j}|$ and eliminate the term
$-J_{i,j}\,x_i\,x_j$ from $H$.

\begin{figure}
\vskip 1.2in 
\includegraphics{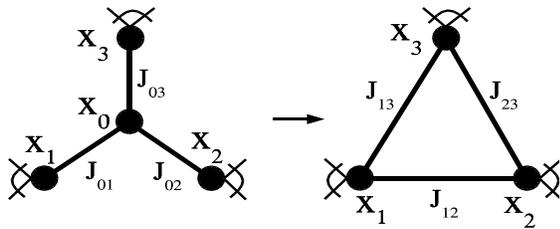}
\caption{``Star-triangle'' relation to reduce a three-connected spin
$x_0$. The new bonds on the right are obtained in
Eq.~(\protect\ref{3coneq}). }
\label{startri}
\end{figure}

\begin{figure}
\vskip 1.7truein \includegraphics{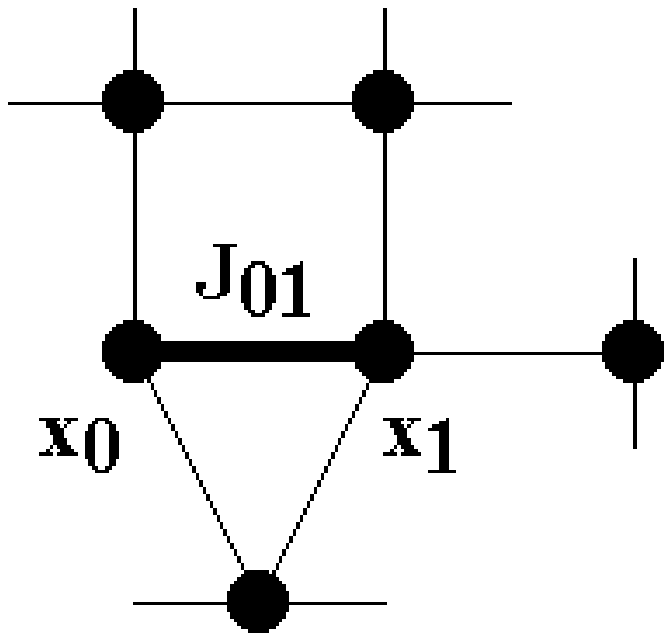} \includegraphics{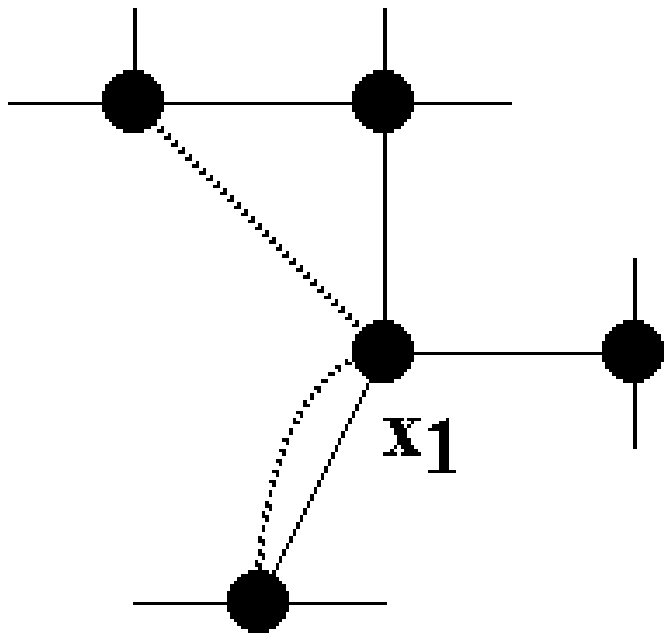}
\caption{Illustration of {\it Rule~VI} for ``strong'' bonds. Left, the
  local topology of a graph is shown for two spins, $x_0$ and $x_1$,
  connected by a bond $J_{0,1}$ (thick line). If $J_{0,1}>0$
  (resp. $J_{0,1}<0$) satisfies Eq.~(\protect\ref{bondeq}), $x_0$ and
  $x_1$ must align (resp. anti-align) in the ground state and $x_0$
  can be removed. Right, the remainder graph is shown after the
  removal. The other bonds emanating from $x_0$ (dashed lines) are now
  directly connected to $x_1$ (with a sign change, if
  $J_{0,1}<0$). This procedure may lead to a double bond ({\it
  Rule~III}), as shown here, if $x_1$ was already connected to a
  neighbor of $x_0$ before.}
\label{superbond}
\end{figure}

\begin{figure*}
\vskip 5.5in 
\includegraphics{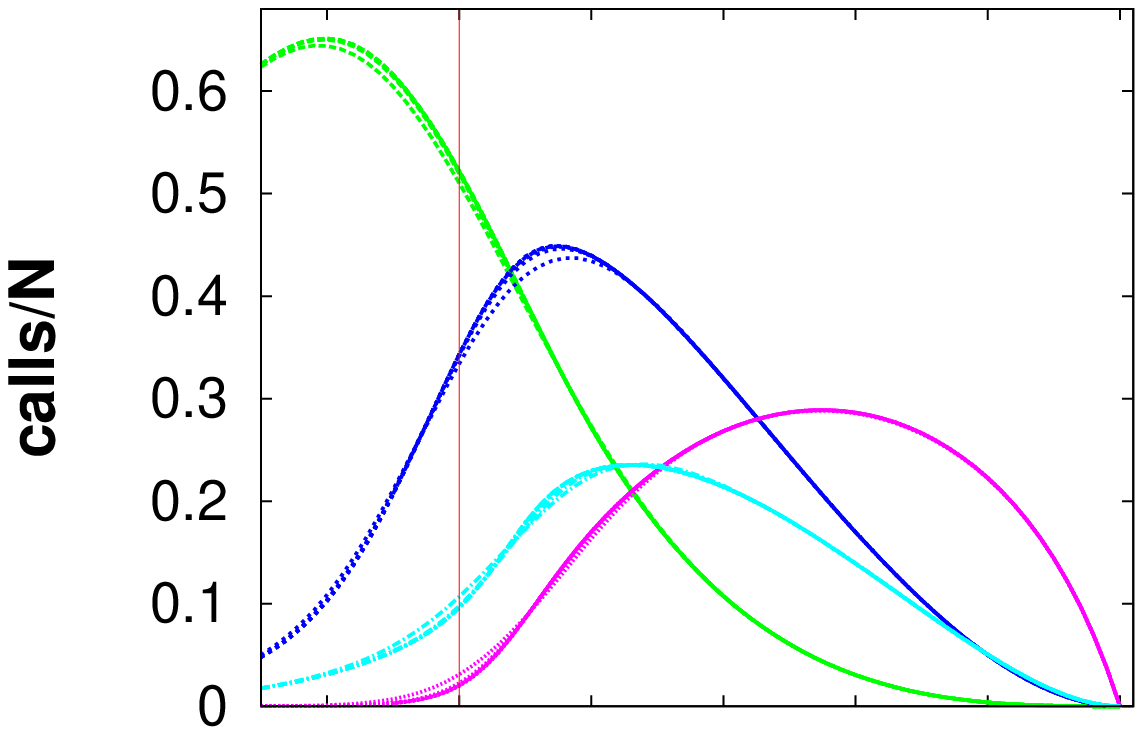}
\includegraphics{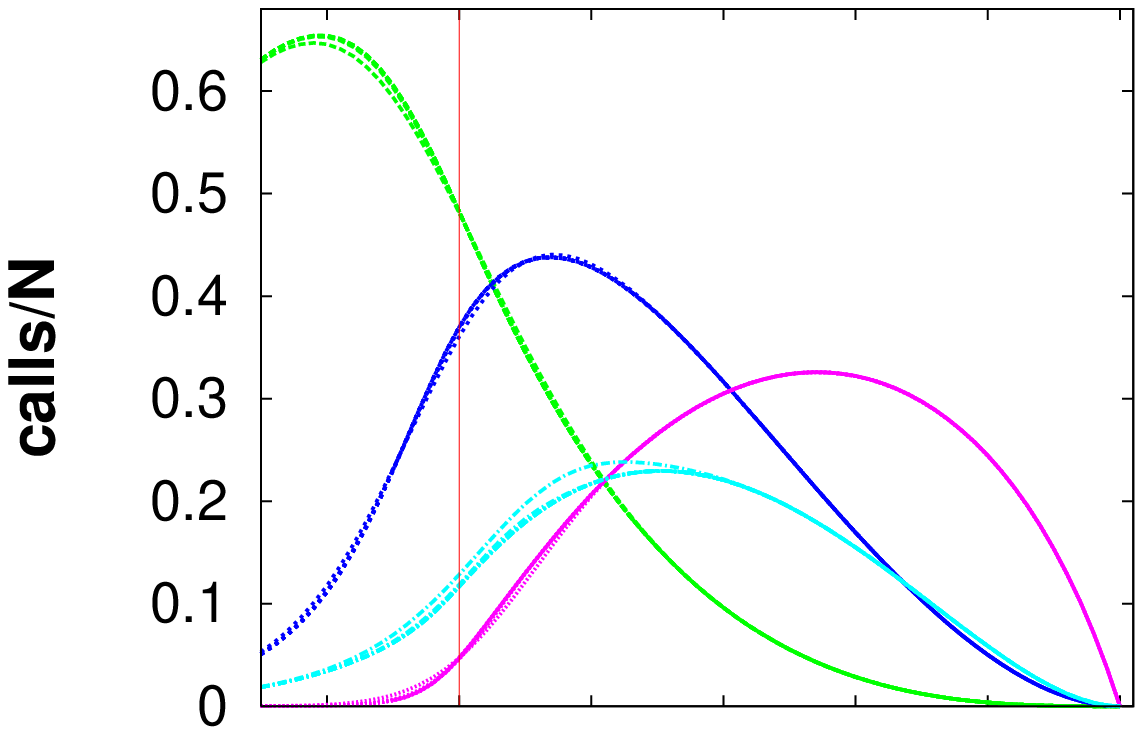}
\includegraphics{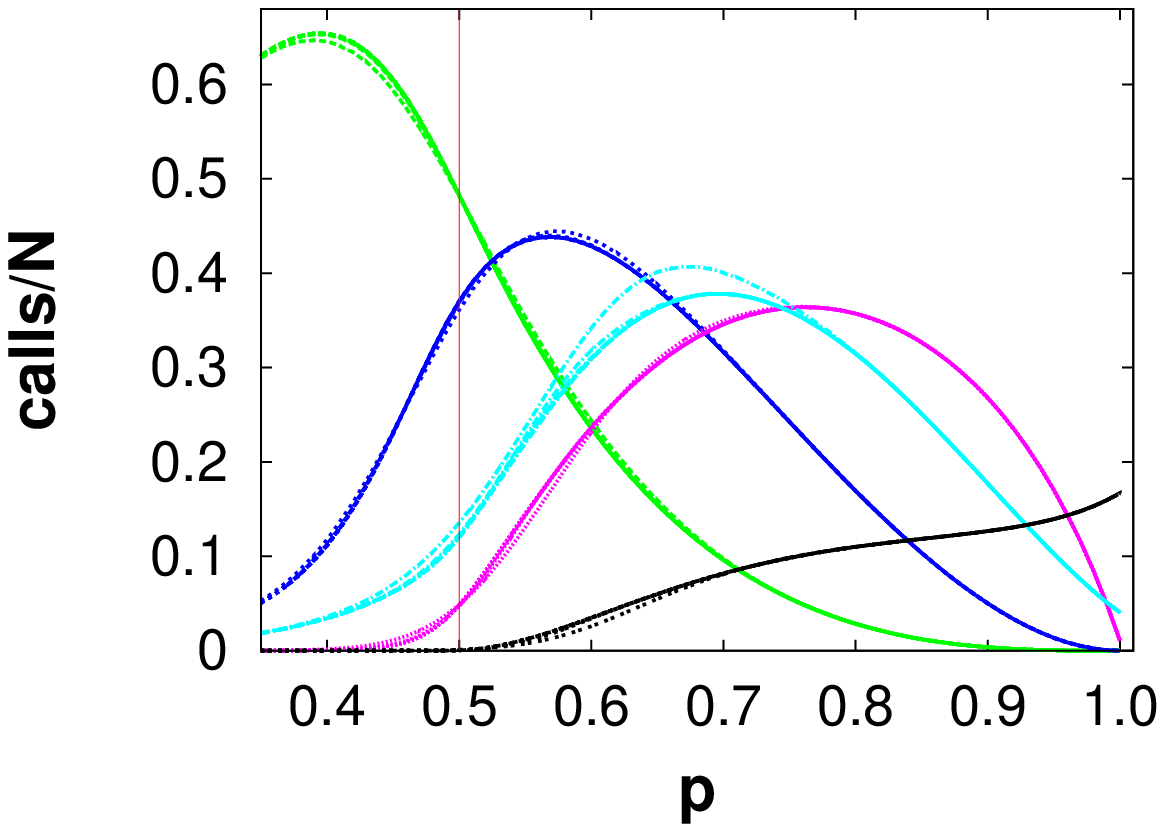}

\includegraphics{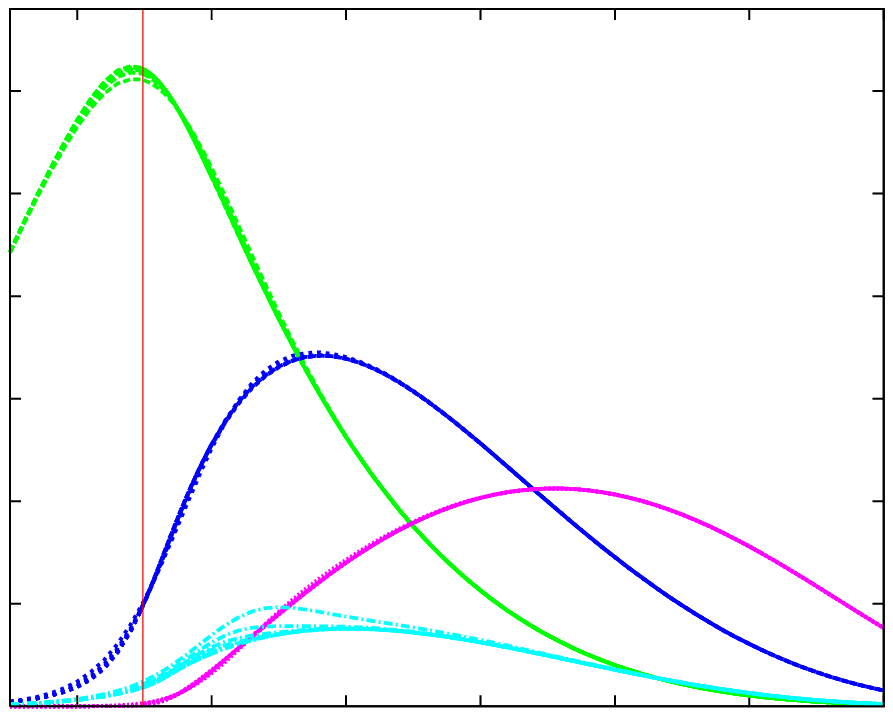}
\includegraphics{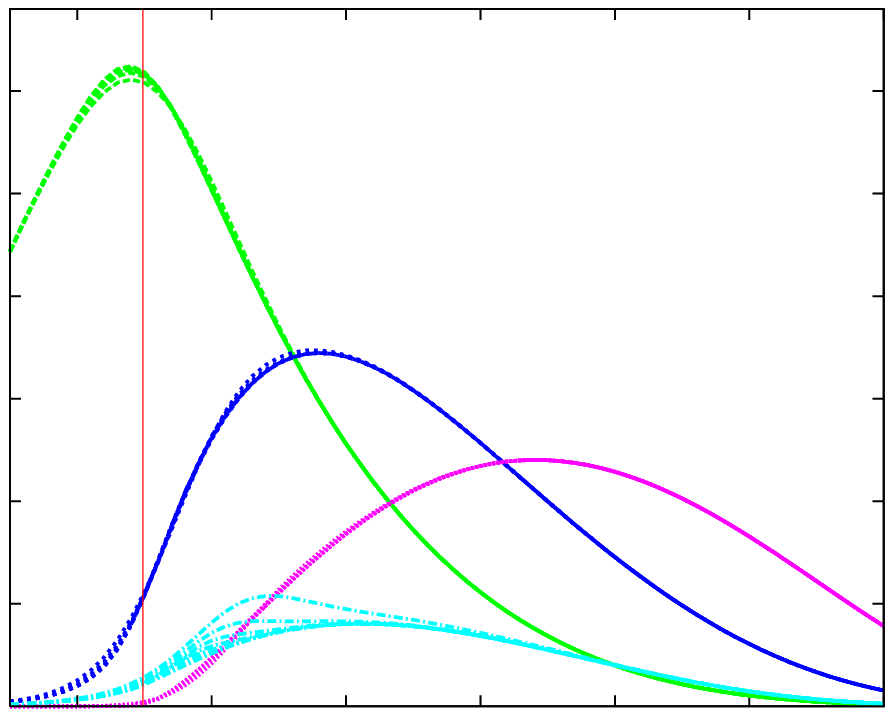}
\includegraphics{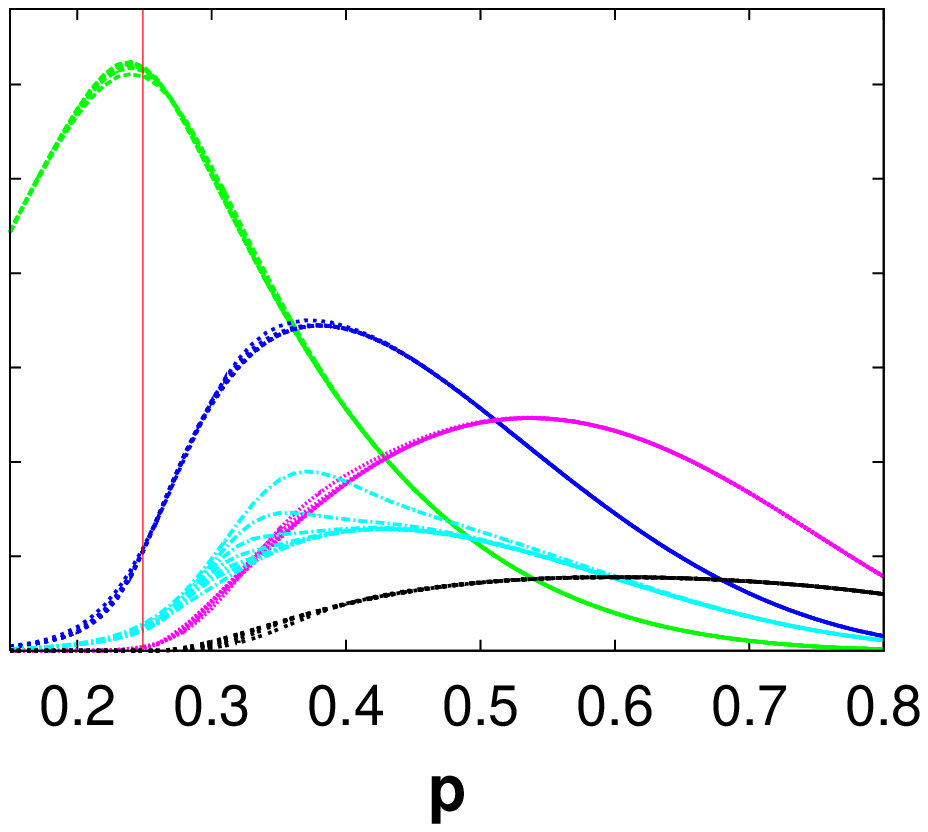}

\includegraphics{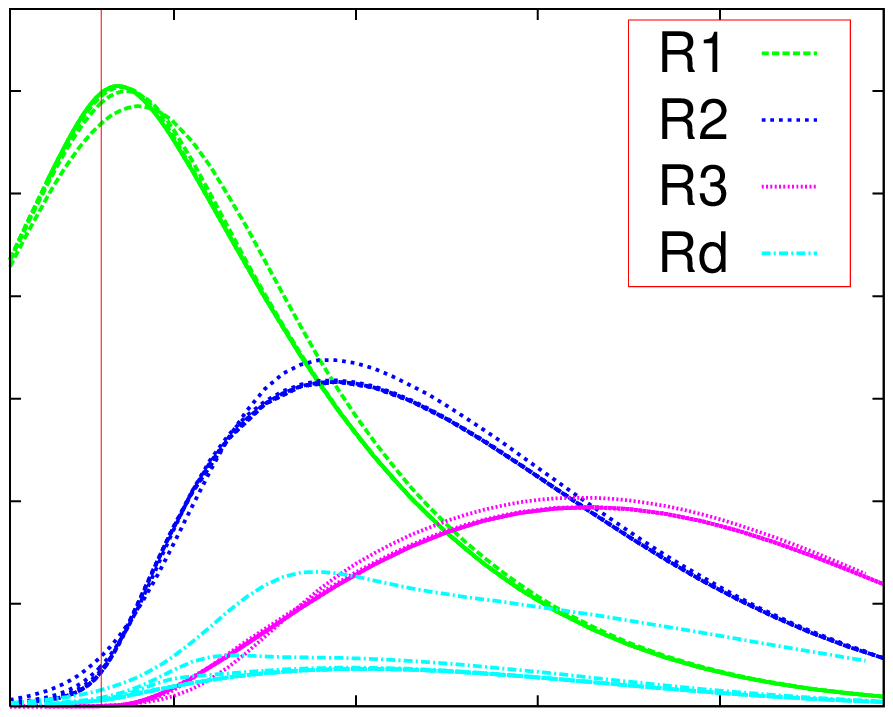}
\includegraphics{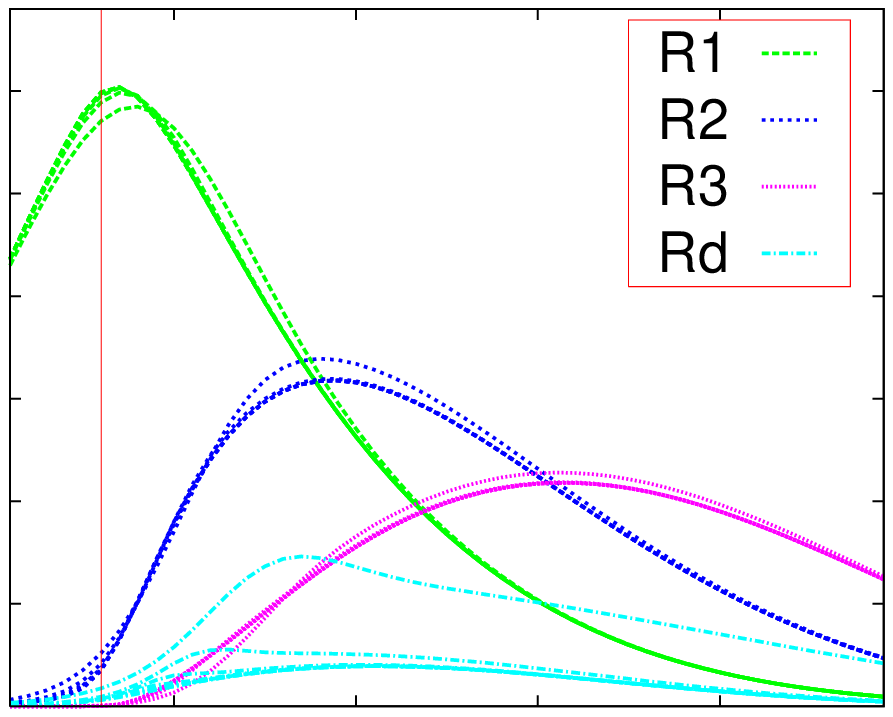}
\includegraphics{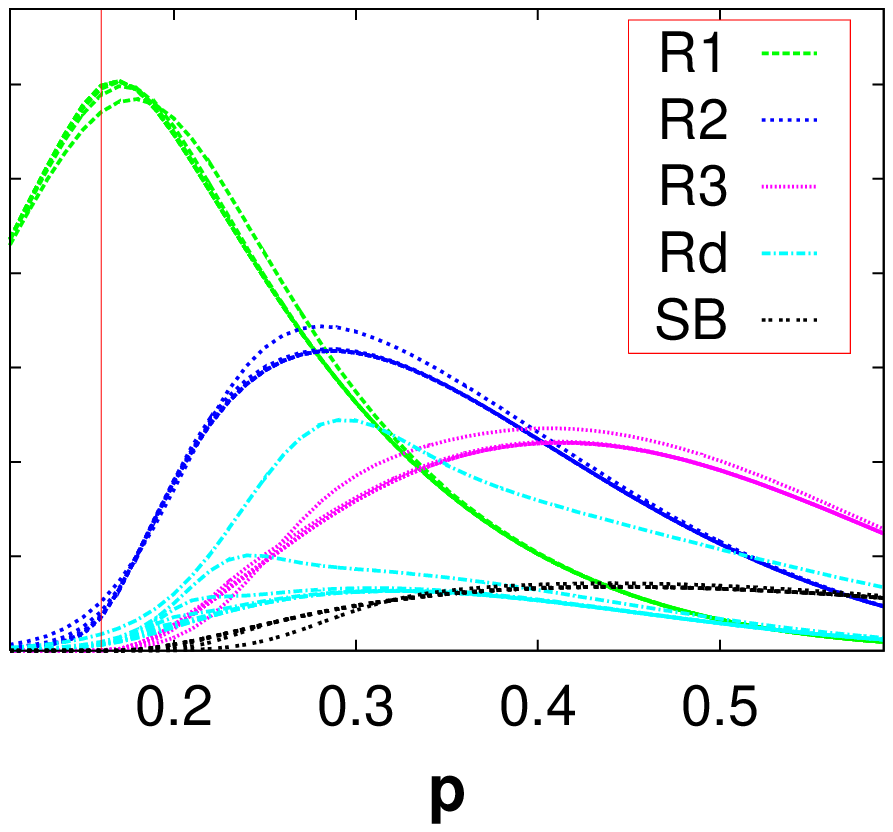}
\caption{Plot of the efficiency of the reduction steps as a function
  of bond density $p$ in $d=2$ (left column), $d=3$ (middle column),
  and $d=4$ (right column) for $\pm J$ bonds (top row), Gaussian bonds
  without super-bond reduction {\bf SB} according to {\it Rule~VI}
  (middle row), and Gaussian bonds with {\bf SB} (bottom row). All
  efficiencies quickly become independent of system size $L$, with
  closer-spaced curves corresponding to larger sizes. {\it
    Rule~VI}, which is useful only for continuous bonds, does not
  effect other rules much, but adds up to 10\% in reduction even at
  $p=1$, with decreasing effect for larger $d$.}
\label{rules_plot}
\end{figure*}

{\it Rule~III:} A double bond, $J_{i,j}^{(1)}$ and $J_{i,j}^{(2)}$,
between two spins $i$ and $j$ can be combined to a single bond by
setting $J_{i,j}= J_{i,j}^{(1)}+J_{i,j}^{(2)}$ or be eliminated
entirely, if the resulting bond vanishes. This operation is very
useful to lower the connectivity of $i$ and $j$ at least by
one. (For an example, see Fig.~\ref{superbond})

{\it Rule~IV:} For a two-connected spin $i$, rewrite the two terms
pertaining to $x_i$ in Eq.~(\ref{Heq}) as
\begin{eqnarray}
x_i(J_{i,1}x_1+J_{i,2}x_2)&\leq&\left|J_{i,1}x_1+J_{i,2}x_2\right|\nonumber\\
&=&J_{1,2}x_1x_2+\Delta H,
\label{2coneq}
\end{eqnarray}
where
\begin{eqnarray}
J_{1,2}&=&\frac{1}{2}\left(\left|J_{i,1}+J_{i,2}\right|-\left|J_{i,1}-J_{i,2}\right|\right),\nonumber\\
\Delta H&=&\frac{1}{2}\left(\left|J_{i,1}+J_{i,2}\right|+\left|J_{i,1}-J_{i,2}\right|\right),
\label{h0eq}
\end{eqnarray}
leaving the graph with a new bond $J_{1,2}$ between spin $1$ and $2$,
and acquiring an offset $H_o:=H_o-\Delta H$. In the ground state, the
bound in Eq.~(\ref{2coneq}) becomes an equality.

{\it Rule~V:} A three-connected spin $i$ can be reduced via a
``star-triangle'' relation, as depicted in Fig.~\ref{startri}. We rewrite the three terms
pertaining to $x_i$ in Eq.~(\ref{Heq}) as:
\begin{eqnarray}
&&J_{i,1}\,x_i\,x_1+J_{i,2}\,x_i\,x_2+J_{i,3}\,x_i\,x_3\nonumber\\
&\leq&\left|J_{i,1}x_1+J_{i,2}x_2+J_{i,3}x_3\right|\label{3coneq}\\
&=&J_{1,2}\,x_1\,x_2+J_{1,3}\,x_1\,x_3+J_{2,3}\,x_2\,x_3+\Delta H,\nonumber
\end{eqnarray}
where
\begin{eqnarray}
&J_{1,2}=-A-B+C+D,\quad J_{1,3}=A-B+C-D,&\nonumber\\
&J_{2,3}=-A+B+C-D,\quad \Delta H=A+B+C+D,&\nonumber\\
&A=\frac{1}{4}\left|J_{i,1}-J_{i,2}+J_{i,3}\right|, \quad
 B=\frac{1}{4}\left|J_{i,1}-J_{i,2}-J_{i,3}\right|,&\nonumber\\
&C=\frac{1}{4}\left|J_{i,1}+J_{i,2}+J_{i,3}\right|,\quad
 D=\frac{1}{4}\left|J_{i,1}+J_{i,2}-J_{i,3}\right|.&\nonumber
\end{eqnarray}
As before, in the ground state, the bound in Eq.~(\ref{3coneq})
becomes an equality.

{\it Rule~VI:} A spin $i$ (of any connectivity) for which the absolute
weight $|J_{i,j'}|$ of one bond to a spin $j'$ is larger than the
absolute sum of all its other bond-weights to neighboring spins
$j\not=j'$, i.~e.
\begin{eqnarray}
|J_{i,j'}|>\sum_{j\not=j'}|J_{i,j}|,
\label{bondeq}
\end{eqnarray}
bond $J_{i,j'}$ {\it must} be satisfied in any ground state. Then,
spin $i$ is determined in the ground state by spin $j'$ and it, as
well as this ``super-bond'' $J_{i,j'}$, can be eliminated accordingly,
as depicted in Fig.~\ref{superbond}. Here, we obtain
$H_0:=H_0-|J_{i,j'}|$. All other bonds connected to $i$ are simply
reconnected with $j'$, but with reversed sign, if $J_{i,j'}<0$.

This procedure is costly, and hence best applied after the other rules
are exhausted. But it can be highly effective for very widely
distributed bonds. In particular, since neighboring spins may reduce
in connectivity and become susceptible to the previous rules again, an
avalanche of further reductions may ensue, see Fig.~\ref{superbond}.

After a recursive application of these rules, the original lattice or
graph is either completely reduced (which is almost always the case
below or near $p_c$), in which case $H_o$ provides the exact ground
state energy already, or we are left with a reduced, compact graph in
which no spin has less than four connections, from which one could
obtain the ground state with some optimization procedure, as described
in Refs.~\onlinecite{Boettcher03a,Boettcher04c}. Reducing even
higher-connected spins would lead to new (hyper-)bonds between more
than two spins, unlike Eq.~(\ref{Heq}), as discussed in the Appendix.

\section{Numerical Simulations}
\label{numerics}
In our simulations we have studied EA spin glasses on
hyper-cubic lattices over a range of sizes $L$ in dimensions $d=2$, 3,
and 4 at various bond fractions $p$ for $\pm J$ bonds and Gaussian
bonds. Similar studies could as well have been done on an arbitrary
family of sparse graphs, without restriction. We have applied the
rules described in Sec.~\ref{reduction} recursively, until no further
reductions were possible. We have calculated a number of aspects of this
reduction, such as the number of spins in the remainder graph relative
to the original lattice, the average connectivity in the remainder
graph, and the fraction of systems that is completely reducible
(without remainder), all as a function of bond-density $p$. Similarly,
we have counted along the way how many times each of the reduction
rules has been applied for graphs of a certain size $L$ and bond
fraction $p$. The system sizes used in each figure of this
  section are listed in Tab.~\ref{sizetable} for each dimension.

\begin{table}
\caption{List of the range of system sizes $L$ chosen for each
  dimension $d$ in Figs.~\ref{rules_plot}-\ref{con_plot}.}
\begin{tabular}{c|cc}
\hline\hline
  $d$ &\qquad& $L$  \\
\hline\hline
2     &&  10, $20,\ldots,100$\\
3     &&   5, $10,\ldots,20$ \\
4     &&   3, $4,\ldots,10,$ 15\\
\hline\hline
\end{tabular}
\label{sizetable}
\end{table}

\begin{figure*}
\vskip 5.5in 
\includegraphics{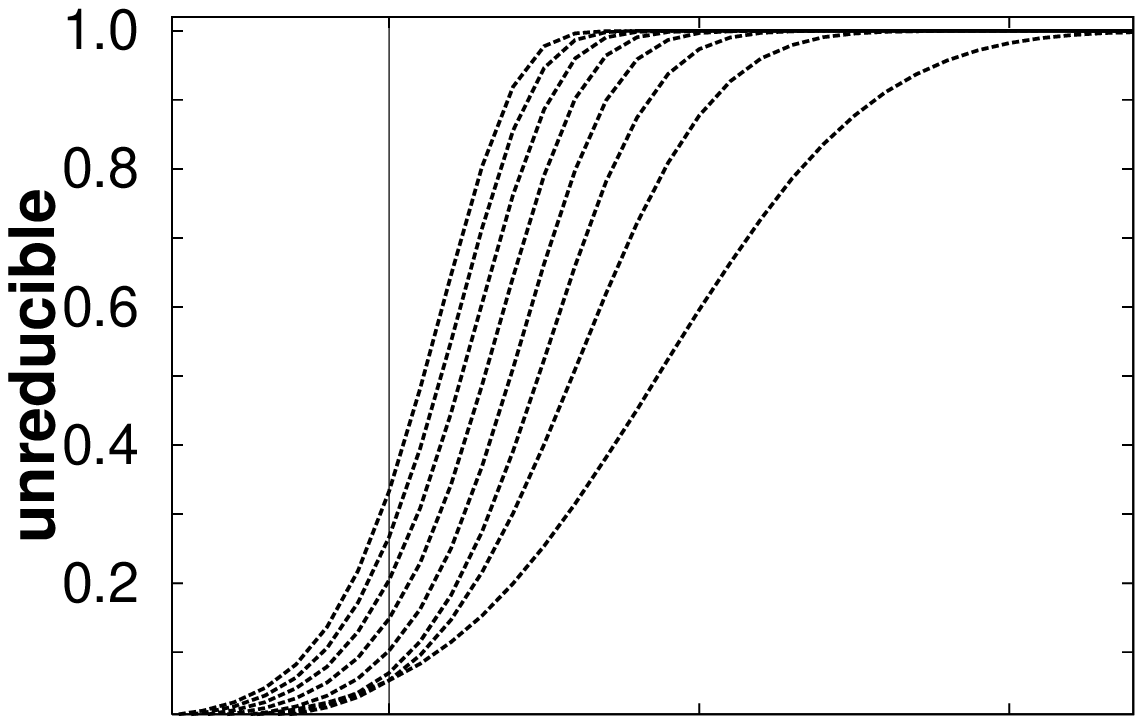}
\includegraphics{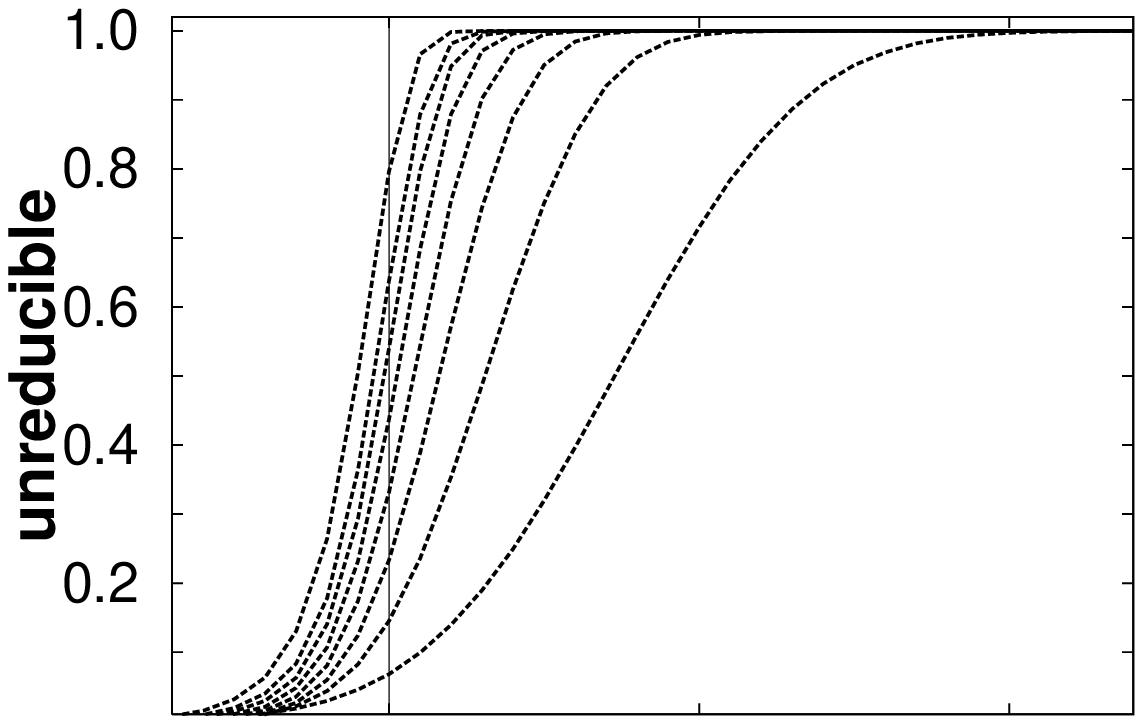}
\includegraphics{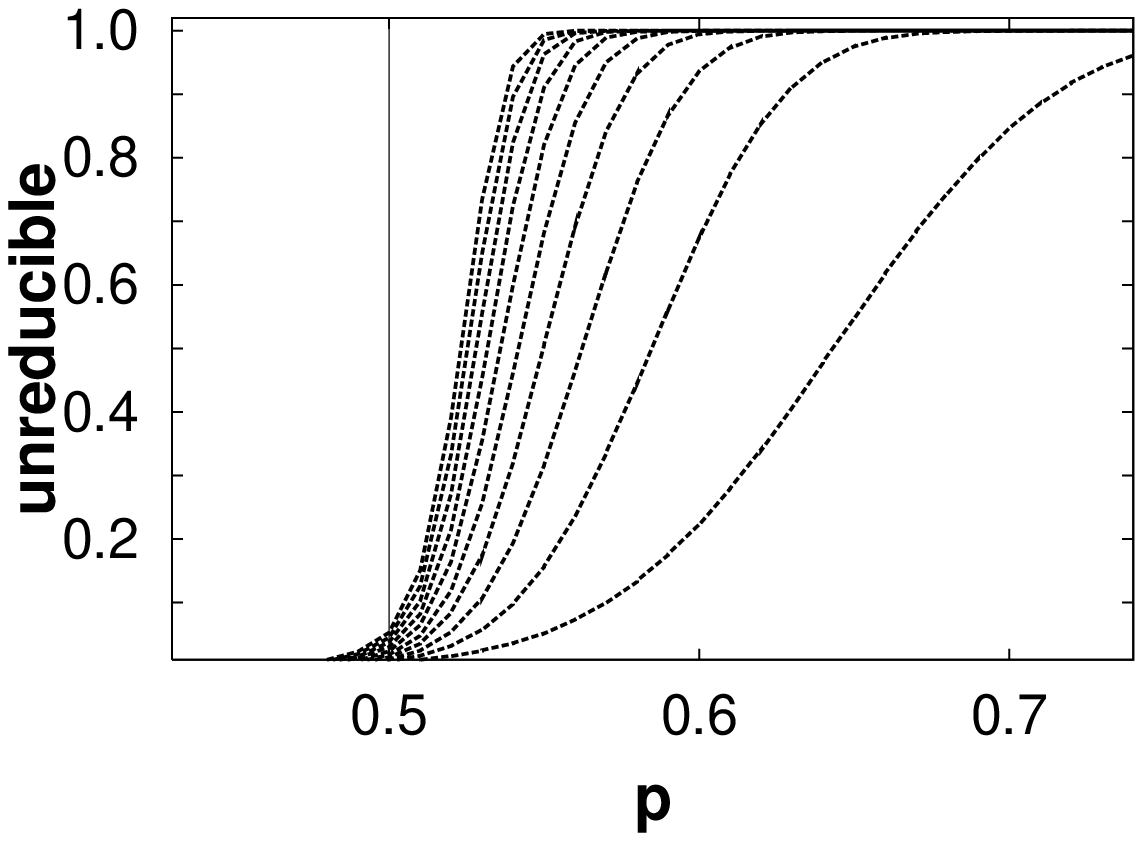}
\includegraphics{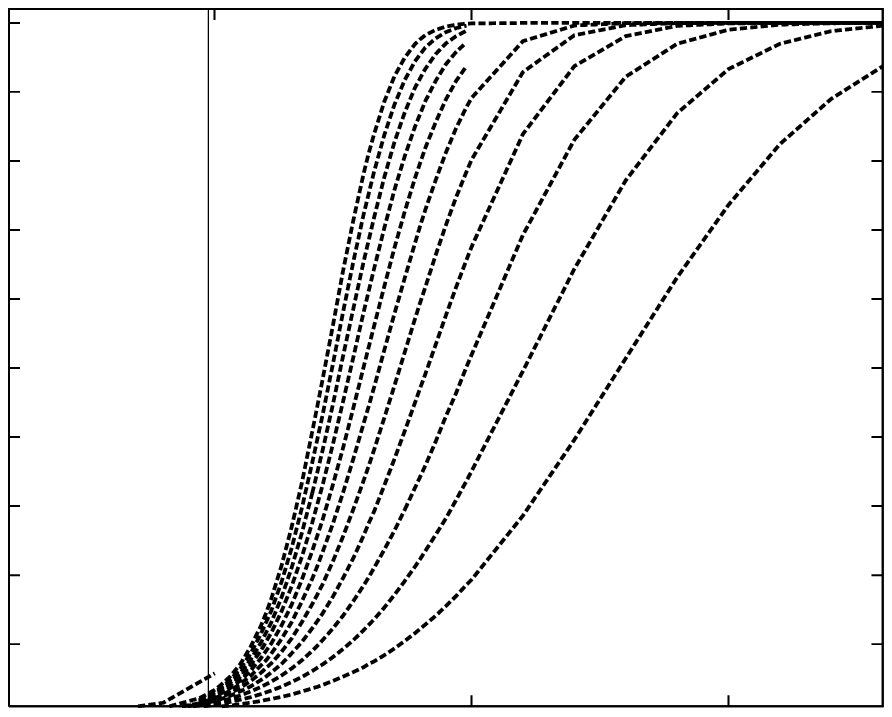}
\includegraphics{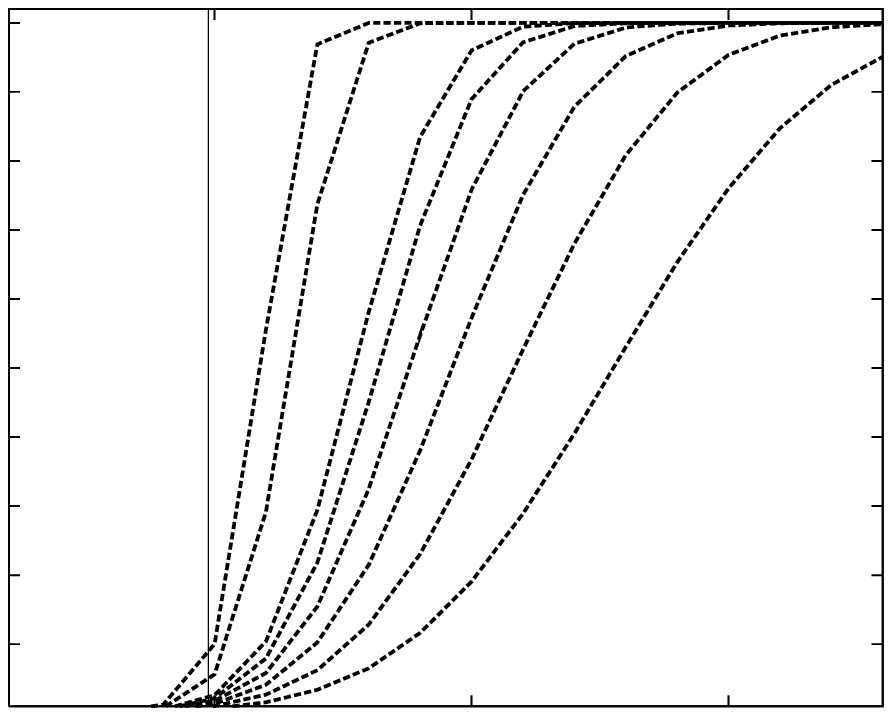}
\includegraphics{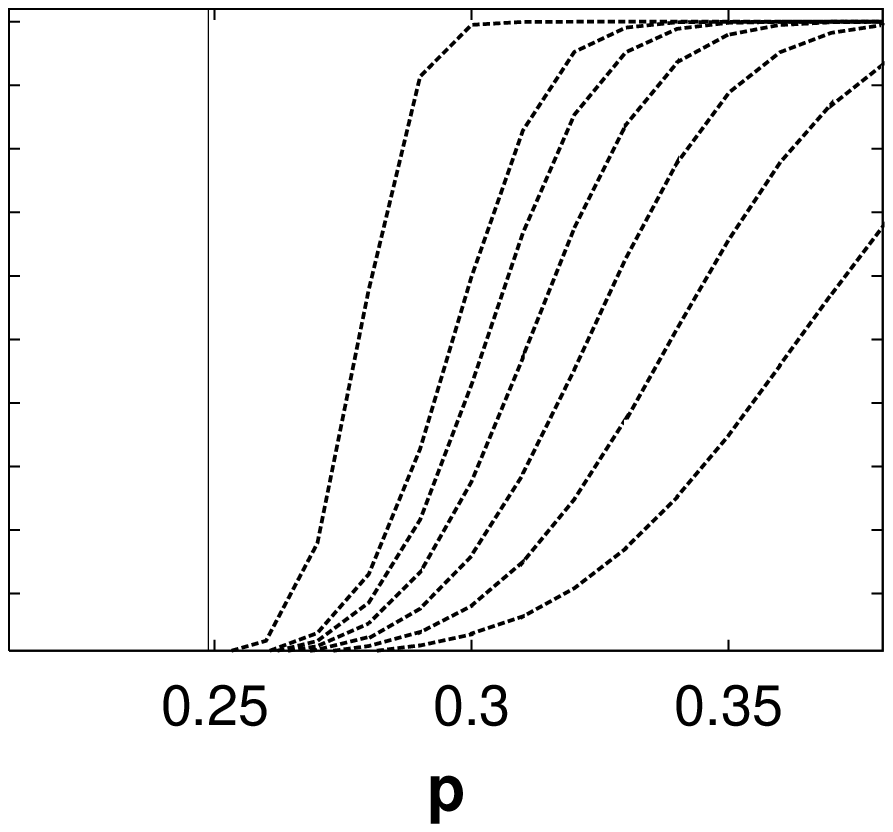}
\includegraphics{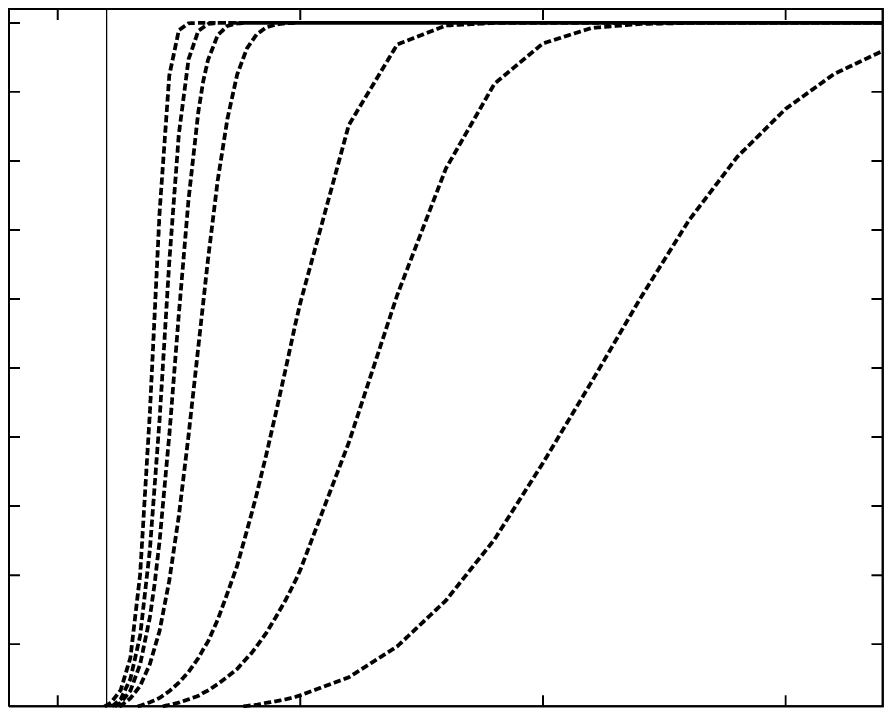}
\includegraphics{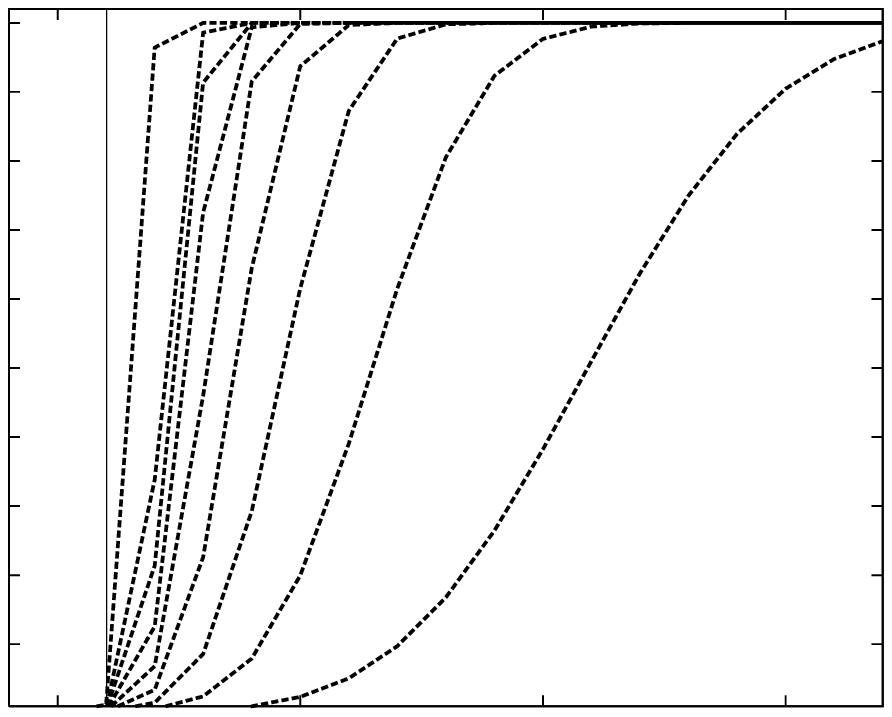}
\includegraphics{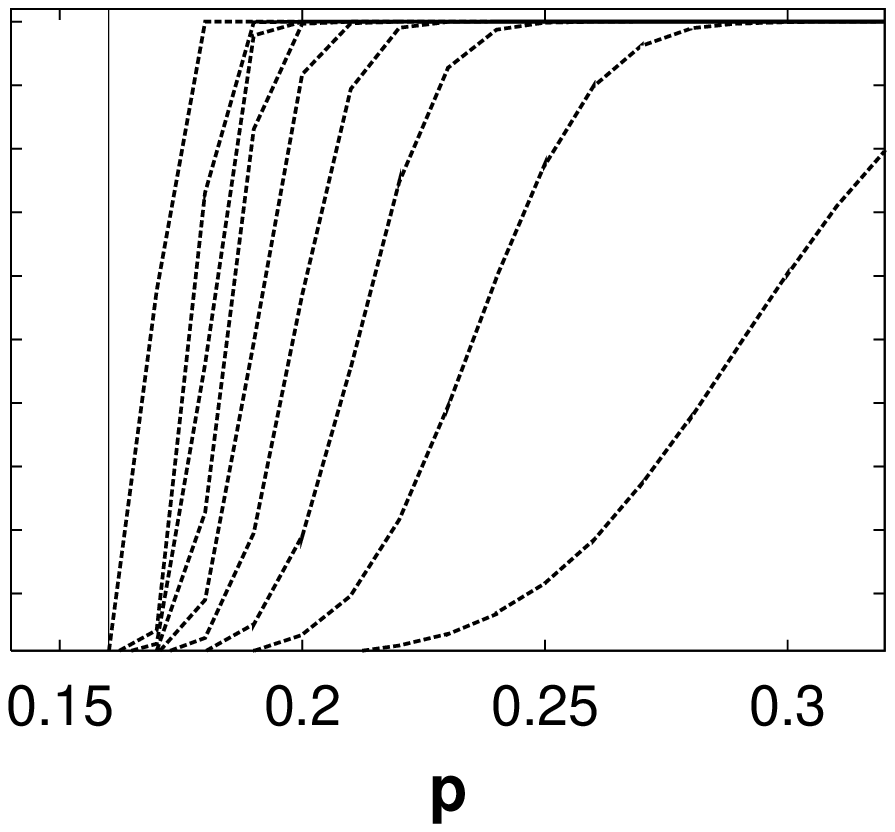}
\caption{Plot of the probability to obtain any non-empty remainder graph after a
  complete exhaustion of the reduction rules as a function of bond
  density $p$ in $d=2$ (left column), $d=3$ (middle column), and $d=4$
  (right column) for $\pm J$ bonds (top row), Gaussian bonds without
  super-bond reduction {\bf SB} according to {\it Rule~VI} (middle
  row), and Gaussian bonds with {\bf SB} (bottom row).  The sequence of
  graphs in each plot steepen for increasing system size $L$ from
  right to left. There is a
  strong dependence on $L$, and it appears that the
  probabilities converge to a $0-1$ step function at or near the bond
  percolation threshold (indicated by a vertical line). With
  super-bond reduction, {\it Rule~VI,} graphs are far more reducible
  even significantly above the threshold, at least at any finite size.}
\label{redux_plot}
\end{figure*}

In Figs.~\ref{rules_plot} we have plotted the efficiency of the
reduction step for one-connected spins ({\it Rule~II} above, labeled
{\bf R1} here), two-connected spins ({\it Rule~IV}, {\bf R2} here),
three-connected spins ({\it Rule~V}, {\bf R3}), double bond
elimination ({\it Rule~III}, {\bf Rd}), and super-bonds ({\it
Rule~VI}, {\bf SB}) as a function of $p$ for dimensions $d=2,$ 3,
and~4. Efficiency is defined here as the number of calls to that step
in a run relative to the original number of spins $N=L^d$ in the
original lattice.  We observe that each of the reduction rules reaches
a peak for increasing bond densities, in order of {\bf R1}, {\bf R2},
{\bf R3}, and {\bf SB}. {\bf Rd}, the elimination of double bonds,
actually does not itself involve the reduction of a spin, and its
behavior is more varied. The rise to that peak is very dependent on
the recursive application of the set of rules, exhausting lower rules
(which are computationally less costly) first before applying a higher
rule. For instance, at least everything that is reducible by {\bf R1}
and {\bf R2} could also have been reduced with {\bf SB}. Thus, the
further to the right a rule peaks, the more powerful it is, and its
decline signals significant changes in the structure of the graph. The
peak of {\bf R1} near $p_c$ (marked by a vertical line in each plot)
is a consequence of the well-known fact that a percolating graph is
predominantly one-connected, i.~e., $p_c\sim1/(2d)$ such that the
connectivity is $\alpha_c=2dp_c\sim1$ for $d\to\infty$. The values for
the bond-percolation thresholds on hyper-cubic lattices
are $p_c=1/2$ in $d=2$, $p_c\approx0.2488$ in $d=3$, and
$p_c\approx0.1601$ in $d=4$. These thresholds are indicated by
vertical lines in each plot.

\begin{figure*}
\vskip 5.5in 
\includegraphics{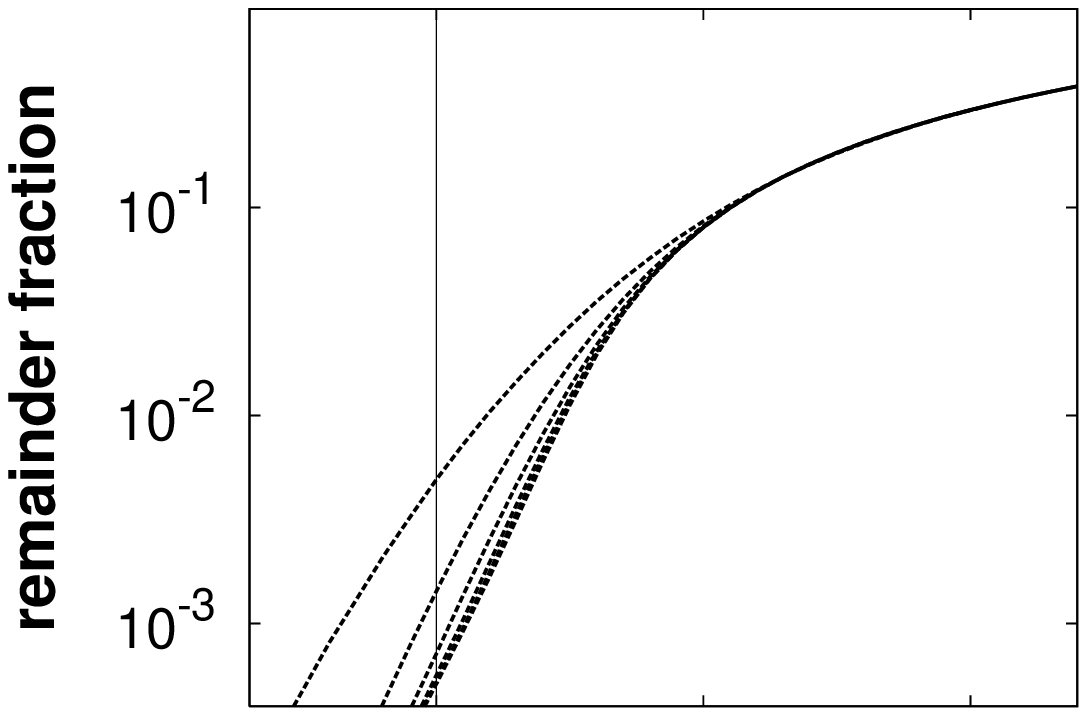}
\includegraphics{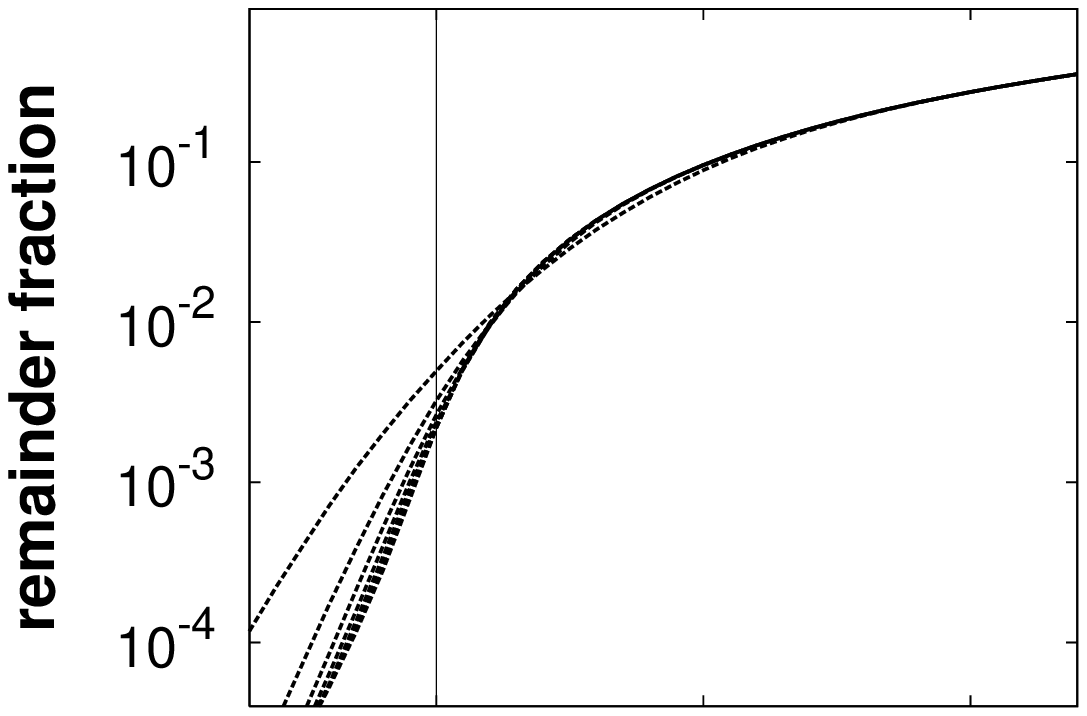}
\includegraphics{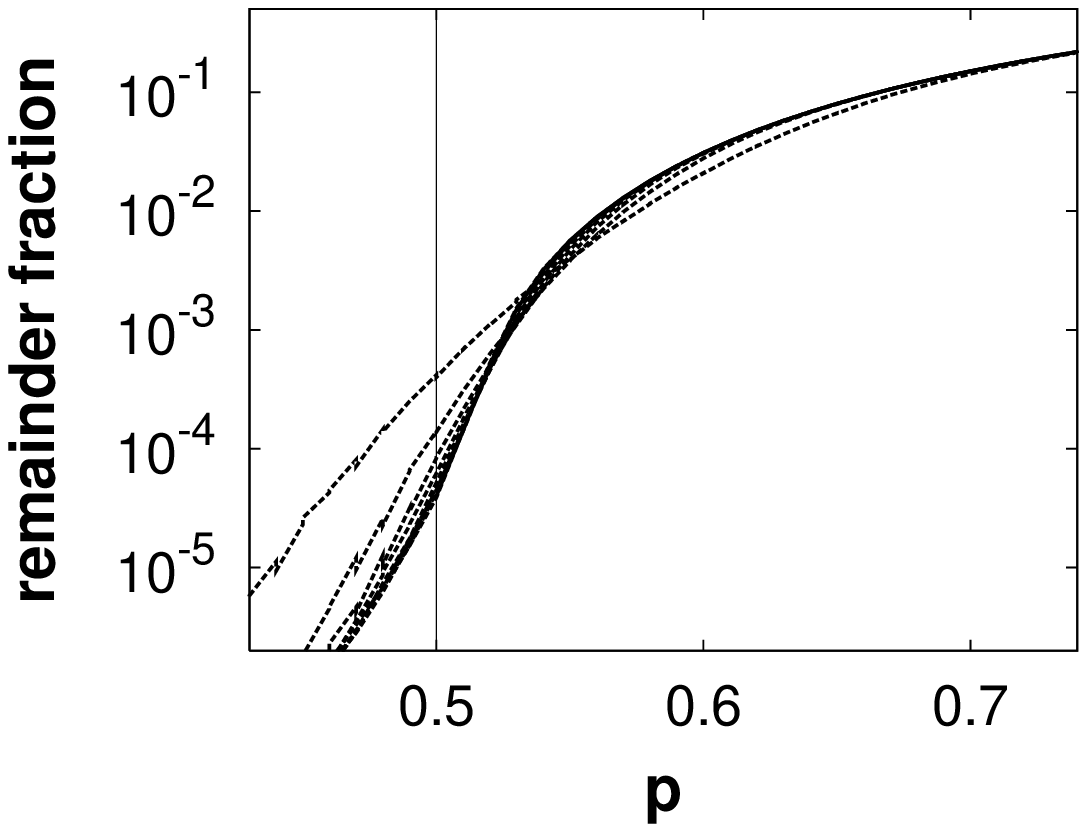}
\includegraphics{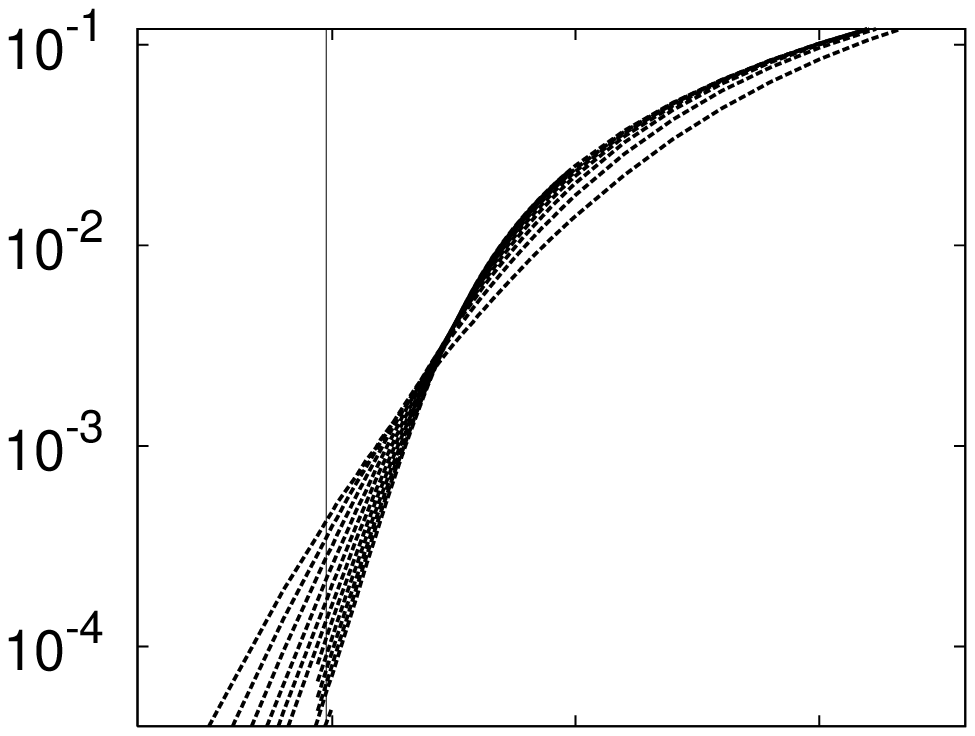}
\includegraphics{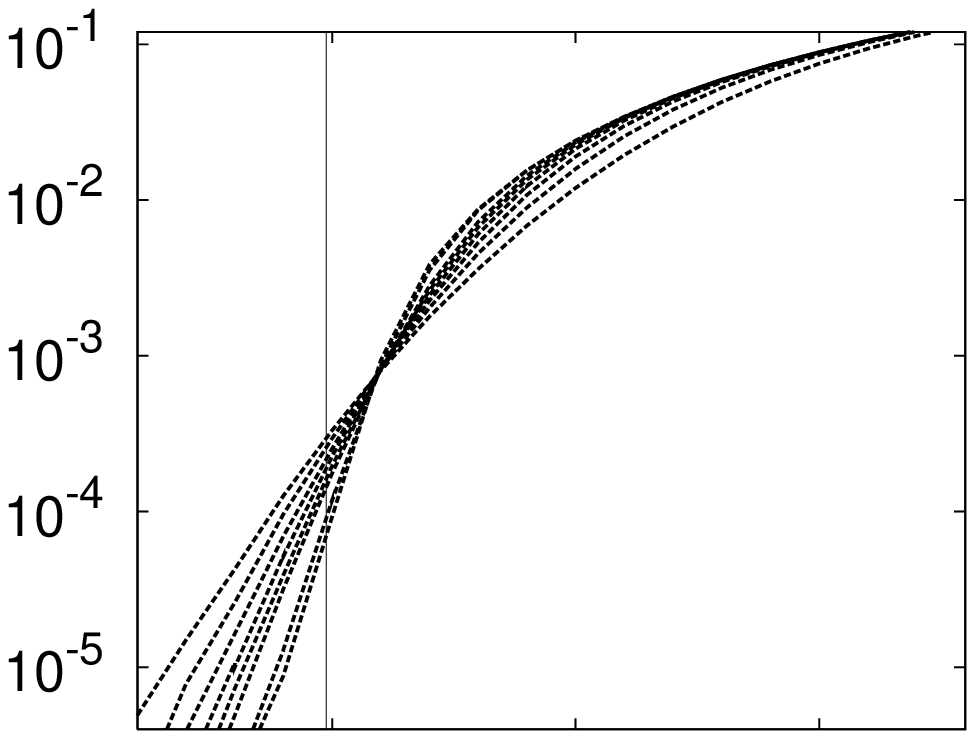}
\includegraphics{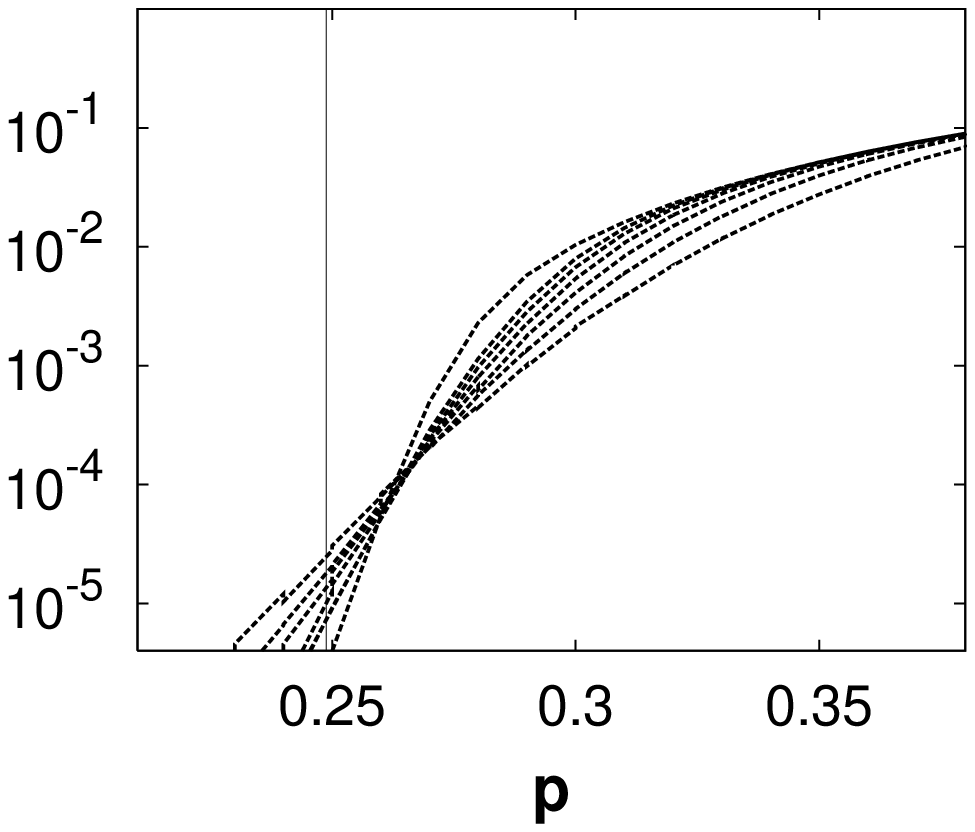}
\includegraphics{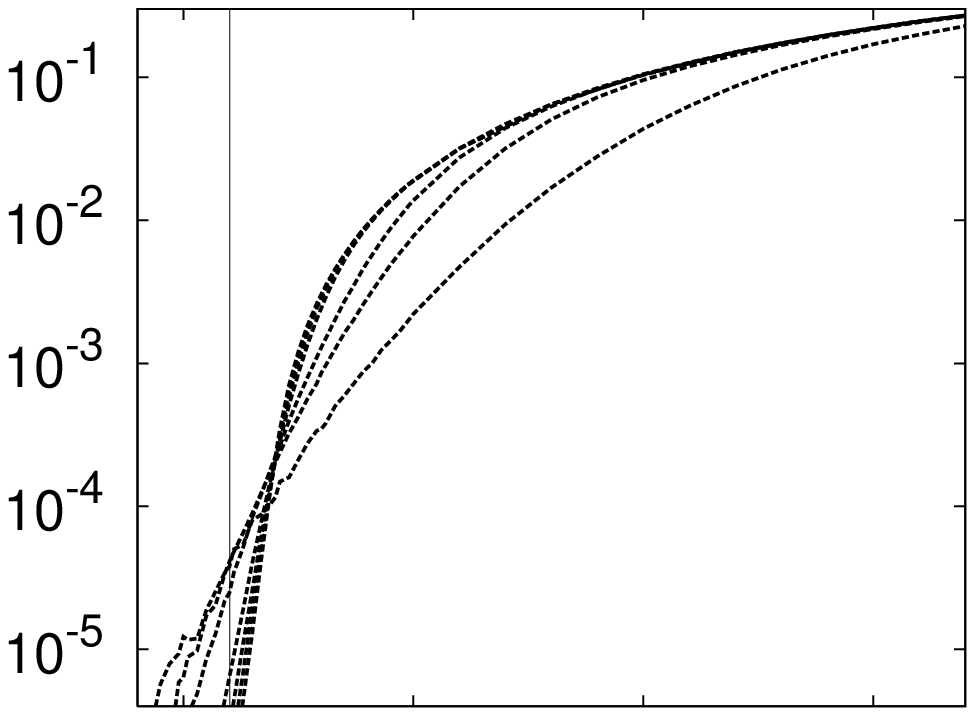}
\includegraphics{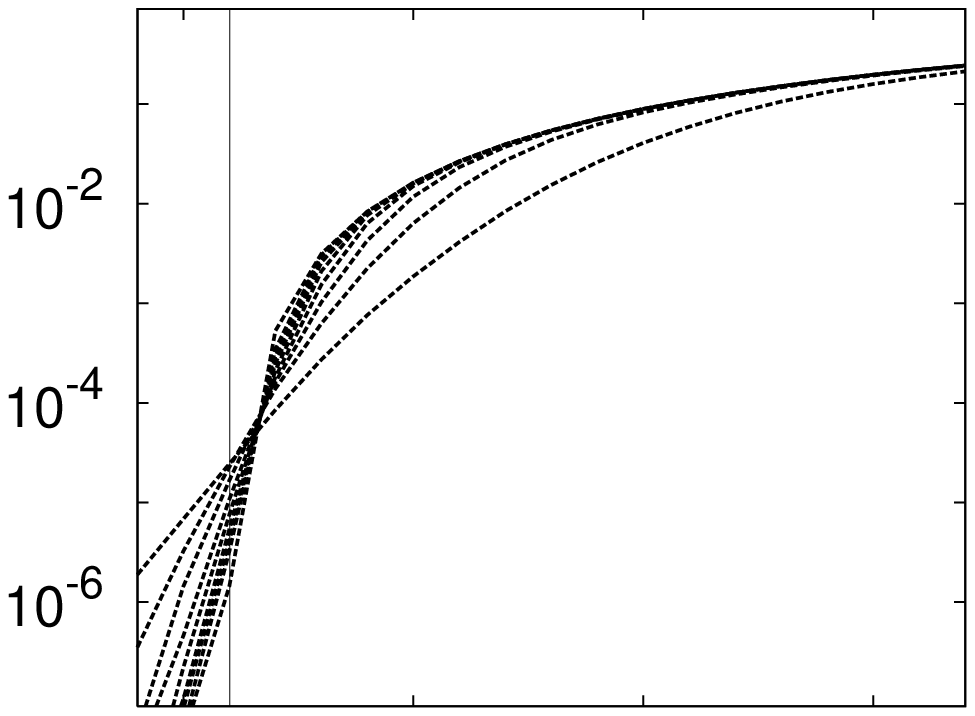}
\includegraphics{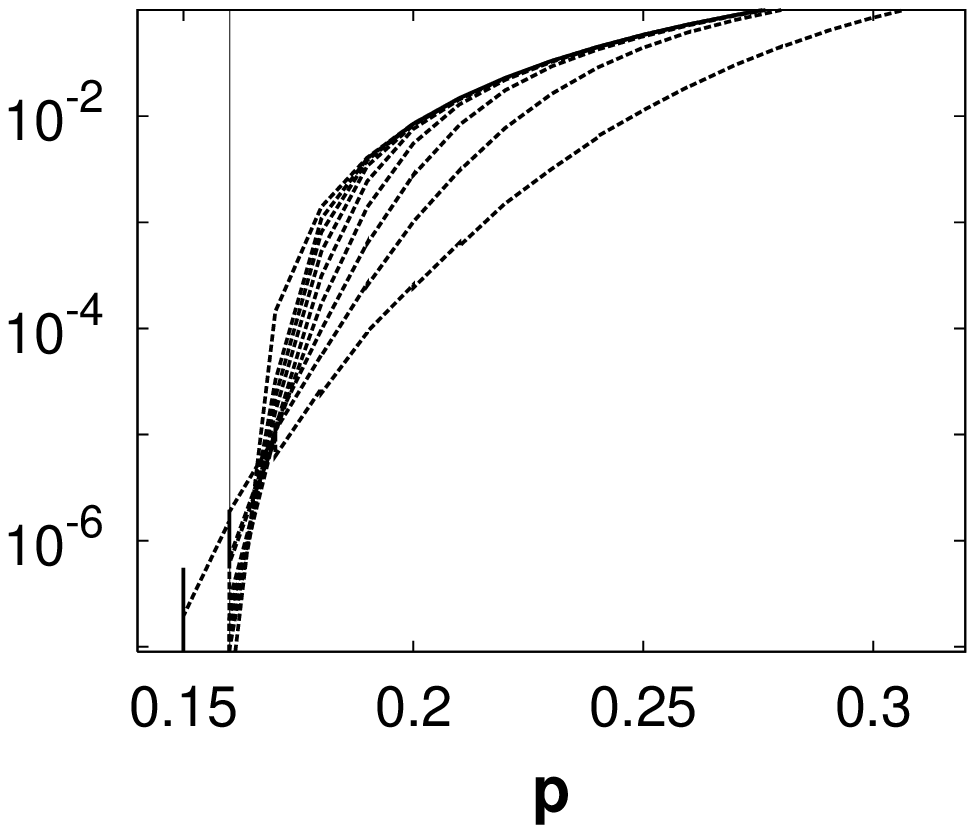}
\caption{Plot of the average (fractional) size of the remainder graph
  (empty or not) as a function of bond density $p$ in $d=2$ (left
  column), $d=3$ (middle column), and $d=4$ (right column) for $\pm J$
  bonds (top row), Gaussian bonds without super-bond reduction {\bf
  SB} according to {\it Rule~VI} (middle row), and Gaussian bonds with
  {\bf SB} (bottom row). Super-bond reduction with {\it Rule~VI}
  lowers the remainder size near the threshold by about an order of
  magnitude, but with diminishing effects for larger $p$. The sequence
  of graphs in each plot steepen for increasing system size $L$.}
\label{remain_plot}
\end{figure*}

The additional use of {\bf SB} does not seem to affect the other rules
much (except for {\bf Rd}). While it does not seem to trigger
avalanches of activity for lower rules (except just above $p_c$), it
in itself often leads to nearly 10\% further reduction at larger $p$.

In Figs.~\ref{redux_plot} we have plotted, as a function of bond
density $p$, the fraction of instances that result in a remainder
graph after a complete exhaustion of the reduction rules. We note that
below and near the bond-percolation threshold $p_c$ in each dimension,
almost all graphs are completely reducible. This implies that the
optimization of their energy can be done in polynomial
time. Physically, this means that there can not be an ordered, glassy
state even at $T=0$ below $p_c$, of course. For increasing system
size, a sharp transition emerges such that almost every graph has some
non-empty remainder (of unspecified size) above that transition. In
case of the discrete $\pm J$ bonds this transition appears to be
related with presumed onset of spin-glass order at $p=p^*>p_c$, as
discussed in Refs.~\onlinecite{Boettcher04c,Boettcher04b}. For Gaussian
bonds, $p^*=p_c$, and the transition appears to be centered close to
that. Asymptotically, the use of {\bf SB} seems to push the transition
just above $p_c$, whereas it seems to locate somewhat below without
{\bf SB}.

In Figs.~\ref{remain_plot} we have plotted the average size of the
remainder graph (empty or not) as a function of bond density
$p$. Including empty remainder graphs in the weight of the average is
important, of course, and explains the values below unity at low $p$.
The pivot point indicates a well-defined transition point closely
related to a 3-core percolation
  transition~\cite{Farrow05,Farrow07,Adler91} above $p_c$, as our
  rules reduce at least all vertices of degree less than 4. The
  correspondence is not exact, as cooperative effects between bond
  weights (such as {\it Rule~III}) or  superbonds ({\it Rule~VI})
  distort the pure case. Predictably, for $p\to1$ the graphs remains
unaltered, except maybe for a few spins reducible by {\bf SB} at lower
$d$.

\begin{figure*}
\vskip 5.5in 
\includegraphics{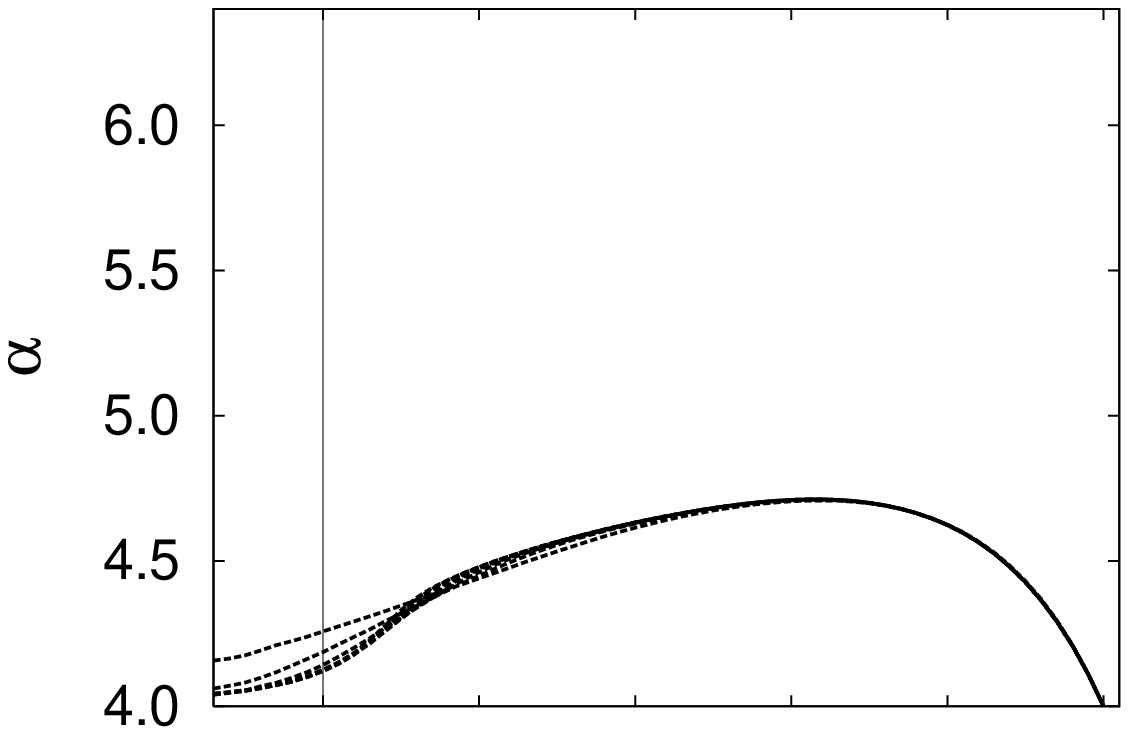}
\includegraphics{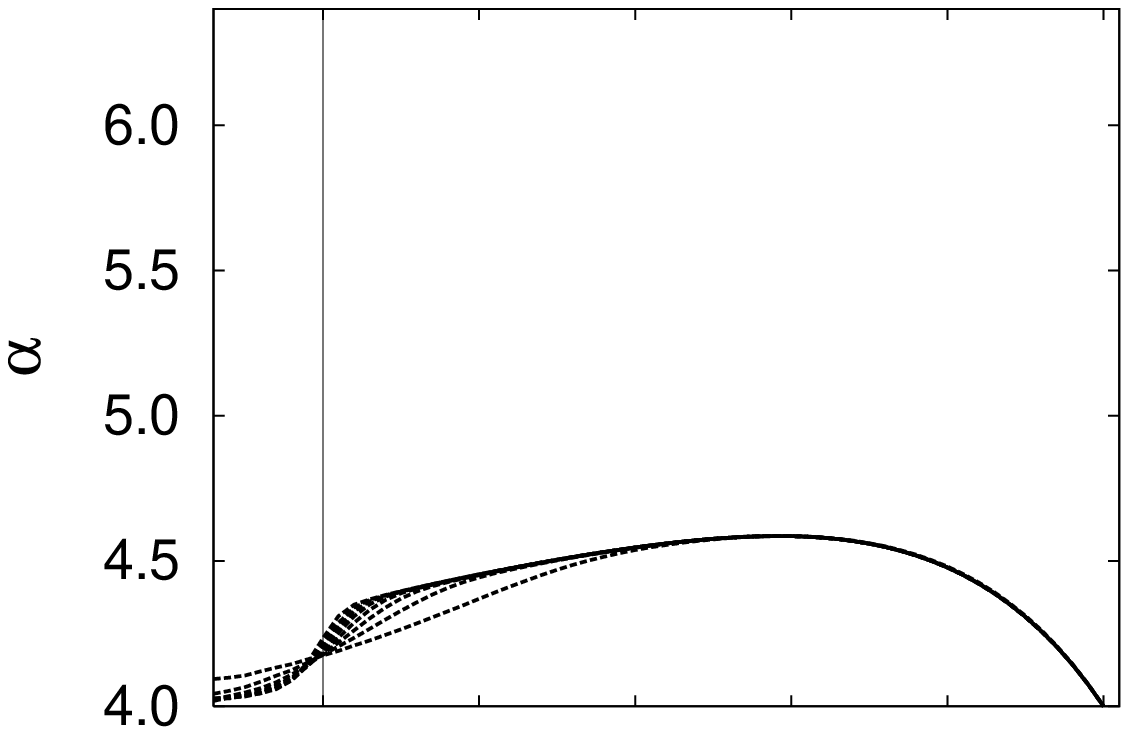}
\includegraphics{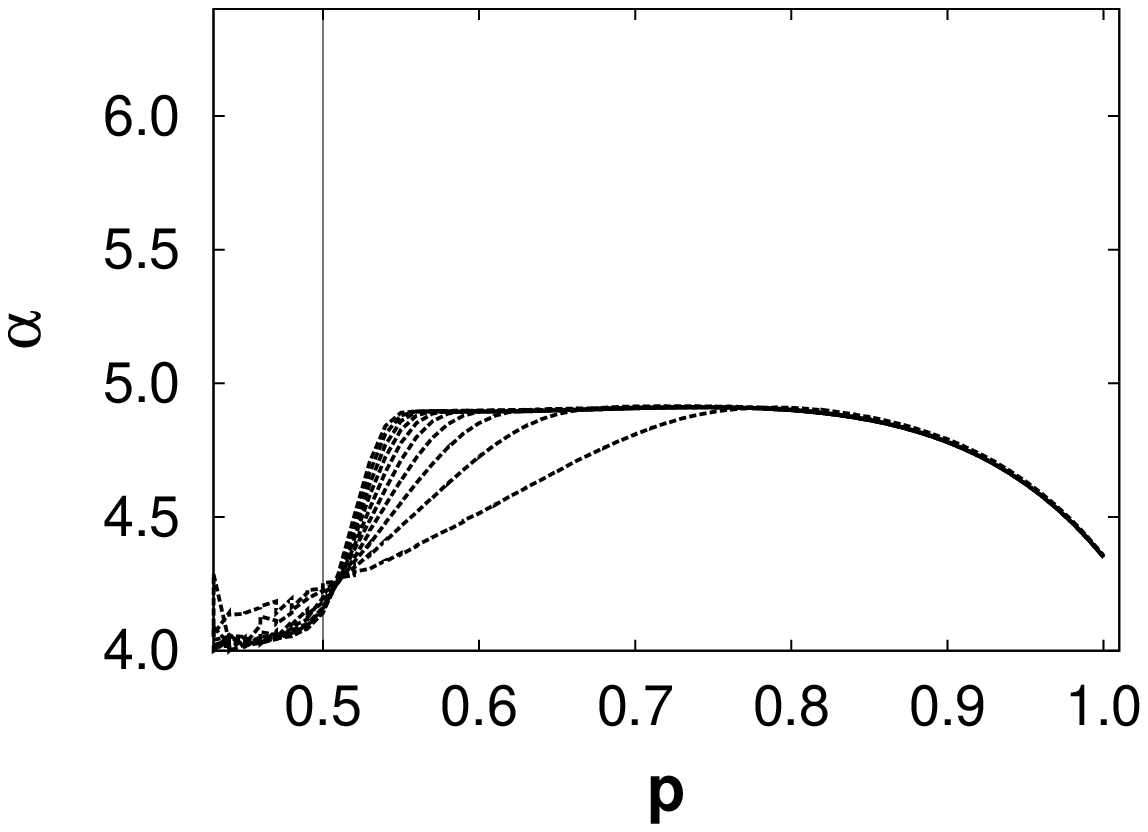}
\includegraphics{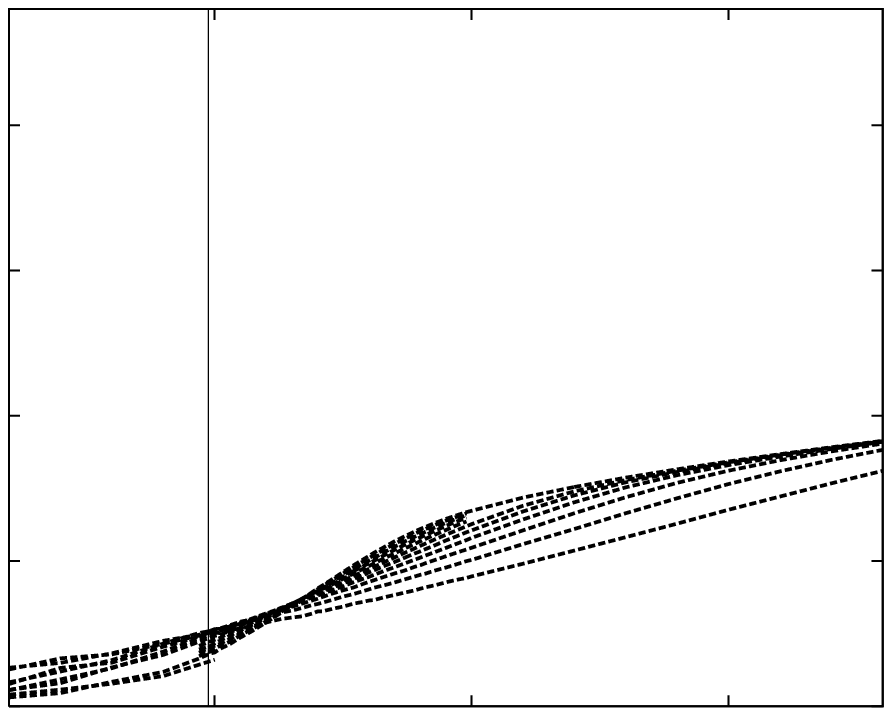}
\includegraphics{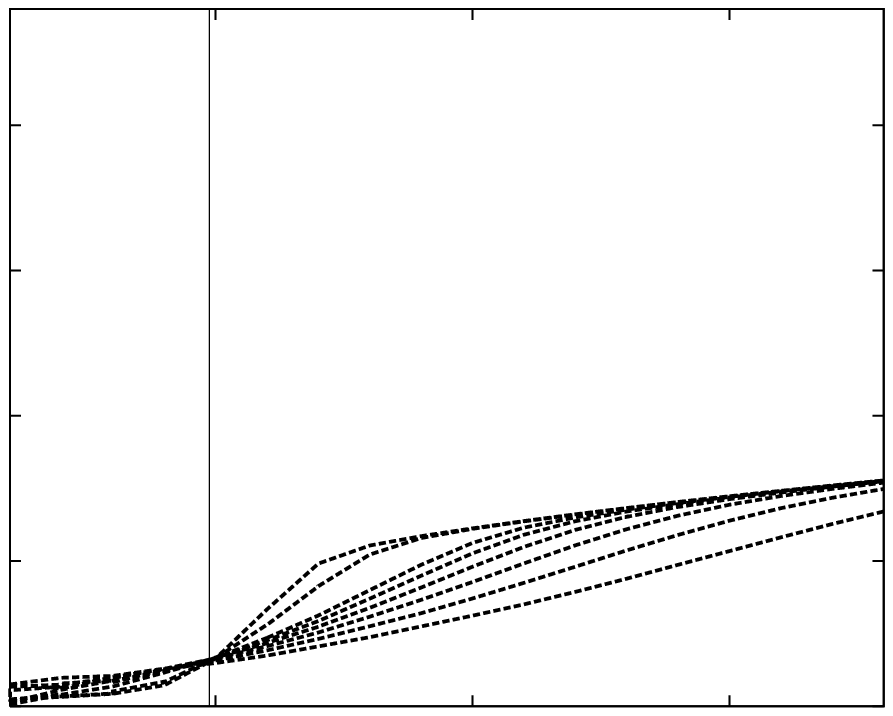}
\includegraphics{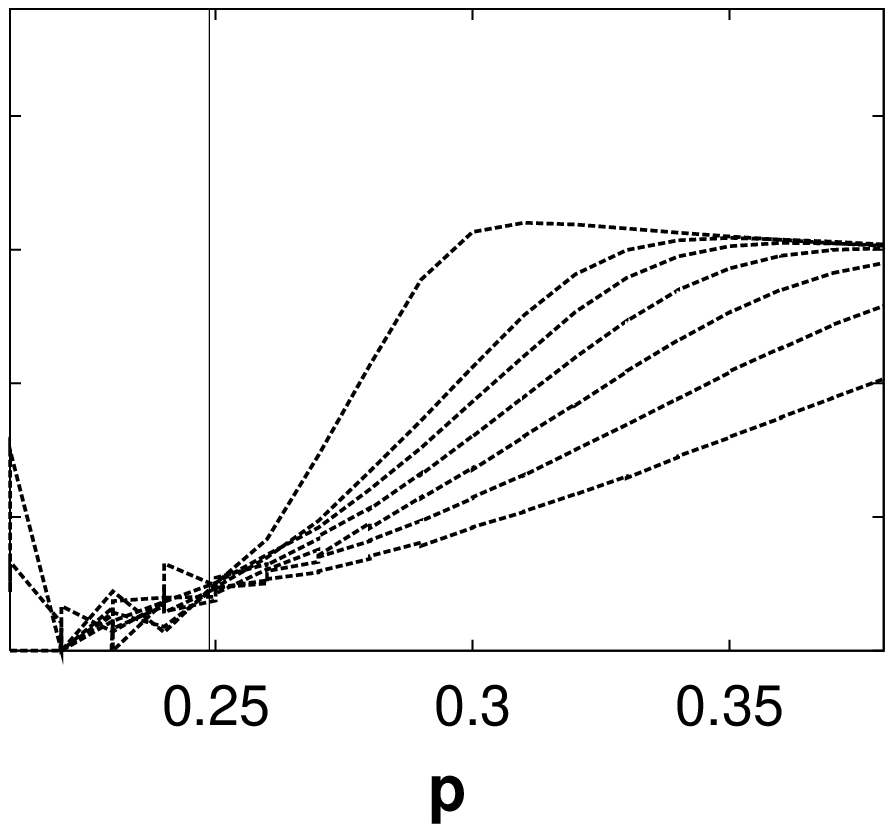}
\includegraphics{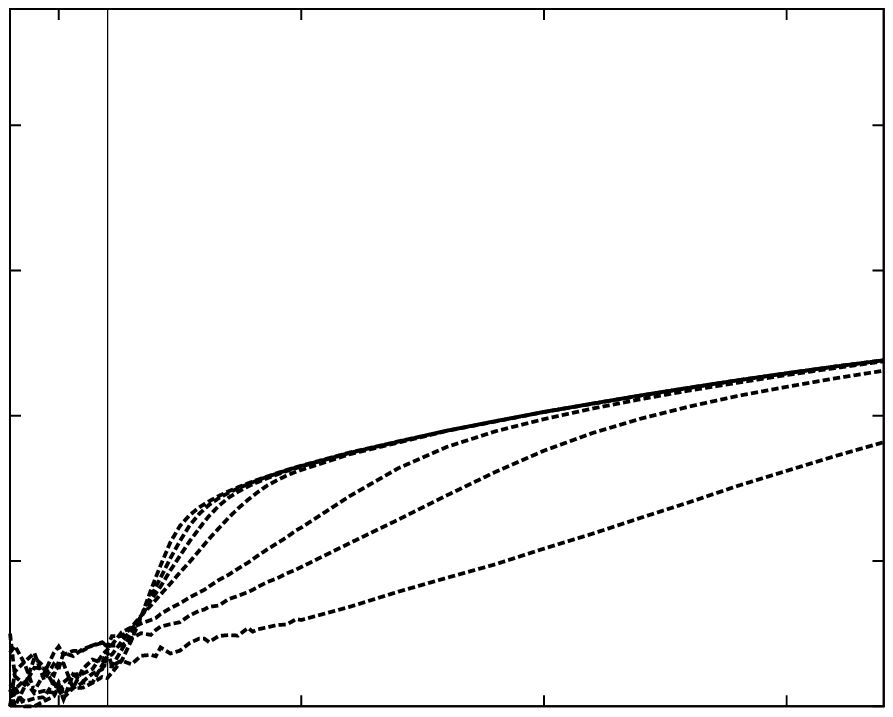}
\includegraphics{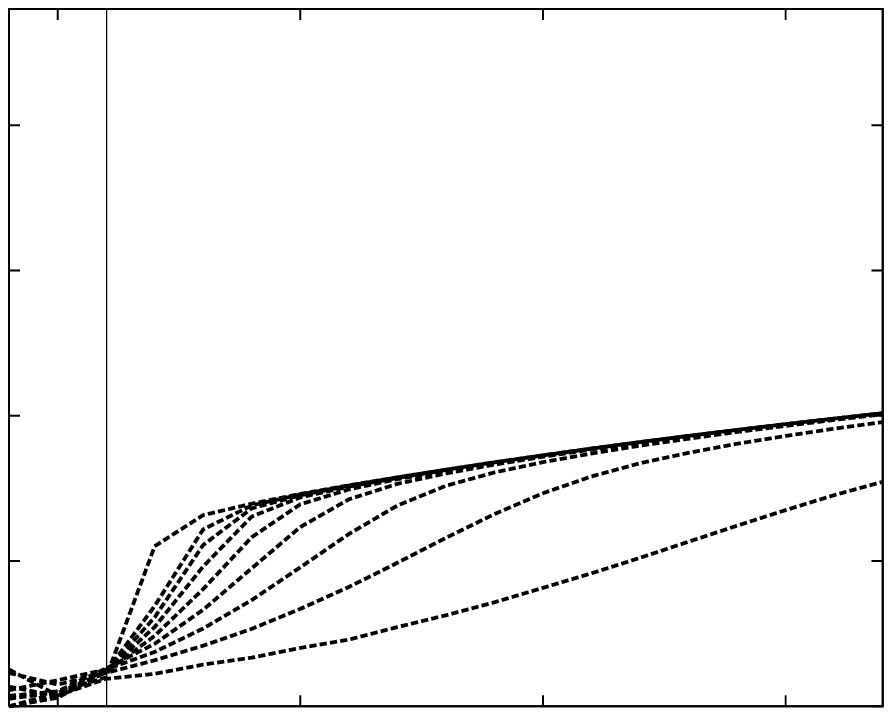}
\includegraphics{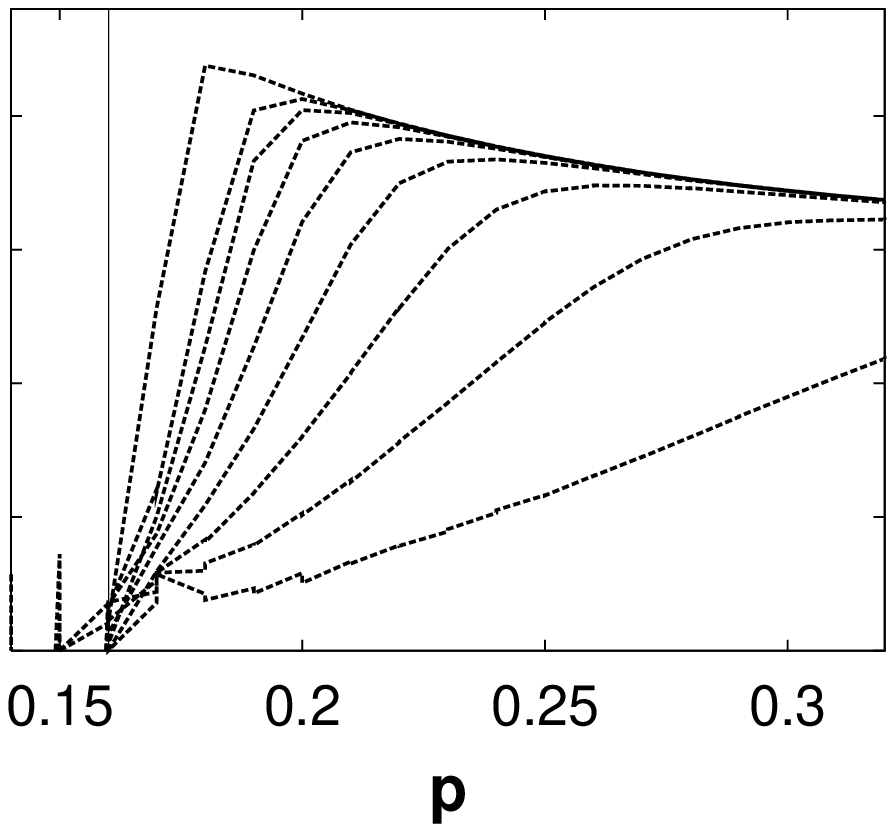}
\caption{Plot of the average connectivity $\langle\alpha\rangle$ of
  any non-empty remainder graph as a function of bond density $p$ in
  $d=2$ (left column), $d=3$ (middle column), and $d=4$ (right column)
  for $\pm J$ bonds (top row), Gaussian bonds without super-bond
  reduction {\bf SB} according to {\it Rule~VI} (middle row), and
  Gaussian bonds with {\bf SB} (bottom row). The dramatic effect of
  implementing {\it Rule~VI} becomes apparent near the
  percolation threshold, especially for increasing $d$. The sequence of
  graphs in each plot steepen for increasing system size $L$.}
\label{con_plot}
\end{figure*}

In Figs.~\ref{con_plot} we have plotted the average connectivity
$\langle\alpha\rangle$ of any non-empty remainder graph as a function
of bond density $p$. By virtue of the reduction rules, it is
$\langle\alpha\rangle\geq4$. The data is very noisy below $p_c$, since
almost all remainders are empty there. These connectivities will
eventually approach $2d$ for $p\to1$, except when {\bf SB} is
included. There is a strong effect due to  {\bf SB} also right above
the threshold $p_c$, where {\it Rule~VI\/} leads to an increasingly
sharper maximum with size and dimension, as Eq.~(\ref{bondeq}) is more
likely satisfied there.

\section{Dominant Bond Reduction Heuristic}
\label{DBR}
The {\it Rule~VI\/} in Sec.~\ref{reduction} is based on the following
observation: If the absolute weight $|J_{i,j'}|$ of one bond incident
on spin $x_i$ from a neighboring spin $x_{j'}$ exceeds the absolute sum of all its other $\alpha_i-1$
bond-weights with adjacent spins, i.~e. if by Eq.~(\ref{bondeq})
\begin{eqnarray}
r_i\equiv\sum_{j=1,j\not=j'}^{\alpha_i}\frac{|J_{i,j}|}{|J_{i,j'}|}<1,
\label{req}
\end{eqnarray}
bond $J_{i,j'}$ {\it must} be satisfied in any ground state. In the
exact reduction procedure, as applied in Sec.~\ref{numerics}, such a
dominant bond is used to eliminate it and the spin $x_i$ from the
problem.

Here, we consider relaxing that constraint to assess the quality of
approximate results that can be obtained with a heuristic approach. We
assume that even if $r_i\geq1$ in Eq.~(\ref{req}), any almost-dominant
bond on a spin $x_i$ is more likely satisfied in a ground state. This
suggest a fast, greedy heuristic: Find the spin $x_i$ with
$r_i=r_{min}=min_{1\leq j\leq N}\{r_j\}$ in an instance and eliminate
it and its strongest bond as in {\it Rule~VI}. This step can be
repeated until any number of the heaviest bonds have been removed to
treat the remainder with an optimization heuristic like EO, or even
until the {\it entire} lattice is reduced. The latter heuristic we
call ``dominant bond reduction'' (DBR).

\begin{figure}
\vskip 2.4truein \includegraphics{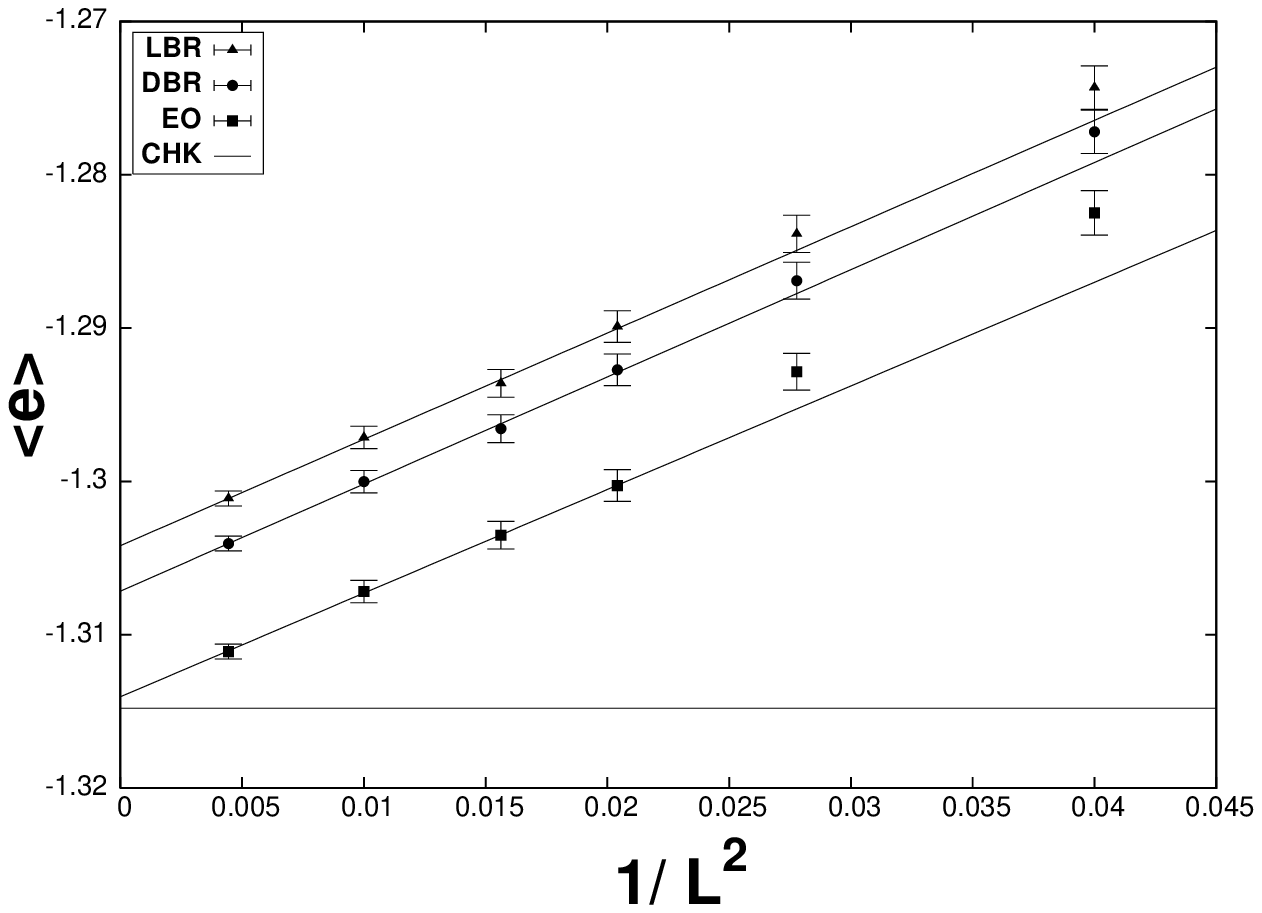}
\caption{Extrapolation plot of approximate ground-state energy
  densities $\langle e\rangle=\langle E\rangle/L^2$ for the
  two-dimensional EA in Eq.~(\ref{Heq}) as a function of $1/L^2$. Each
  data point is the average over $10^4$ random instances. The results
  for the DBR heuristic (circles) discussed in the text extrapolate to
  within 0.5\% of the best-known prediction by Campbell, Hartmann, and
  Katzgraber (CHK)~\cite{Campbell04}, marked by the horizontal
  line. Shown are also results obtained with the EO heuristic
  (squares)~\cite{Boettcher01a,Boettcher00}, which reproduces CHK
  within errors, and a simple alternative to DBR we call Least Bond
  Reduction (LBR, triangles). LBR also applies {\it Rule~VI}, but to
  the spin with the weakest total bond-weight, i.~e., where the sum of
  all absolute weights of incident bonds is minimal. Both, DBR and
  LBR, are much faster than EO. LBR is marginally simpler than DBR and
  avoids the creation of highly-connected spins.  }
\label{2dextra}
\end{figure}

\begin{figure}
\vspace{3.13in}
\includegraphics{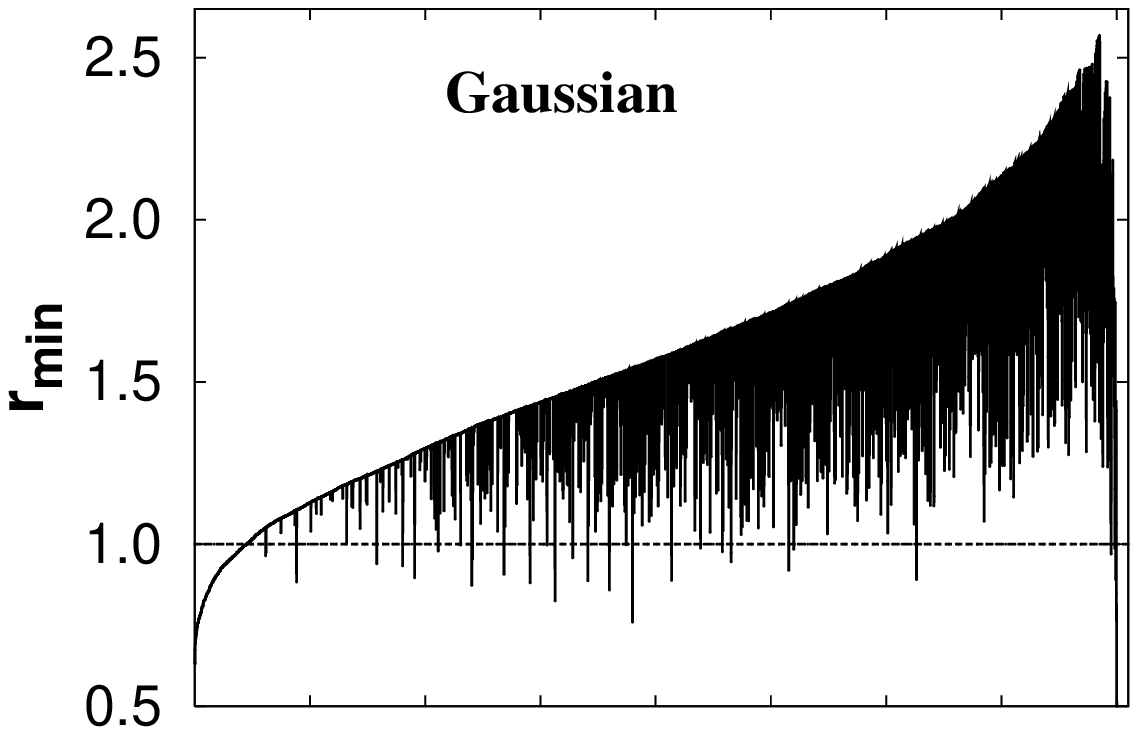}
\includegraphics{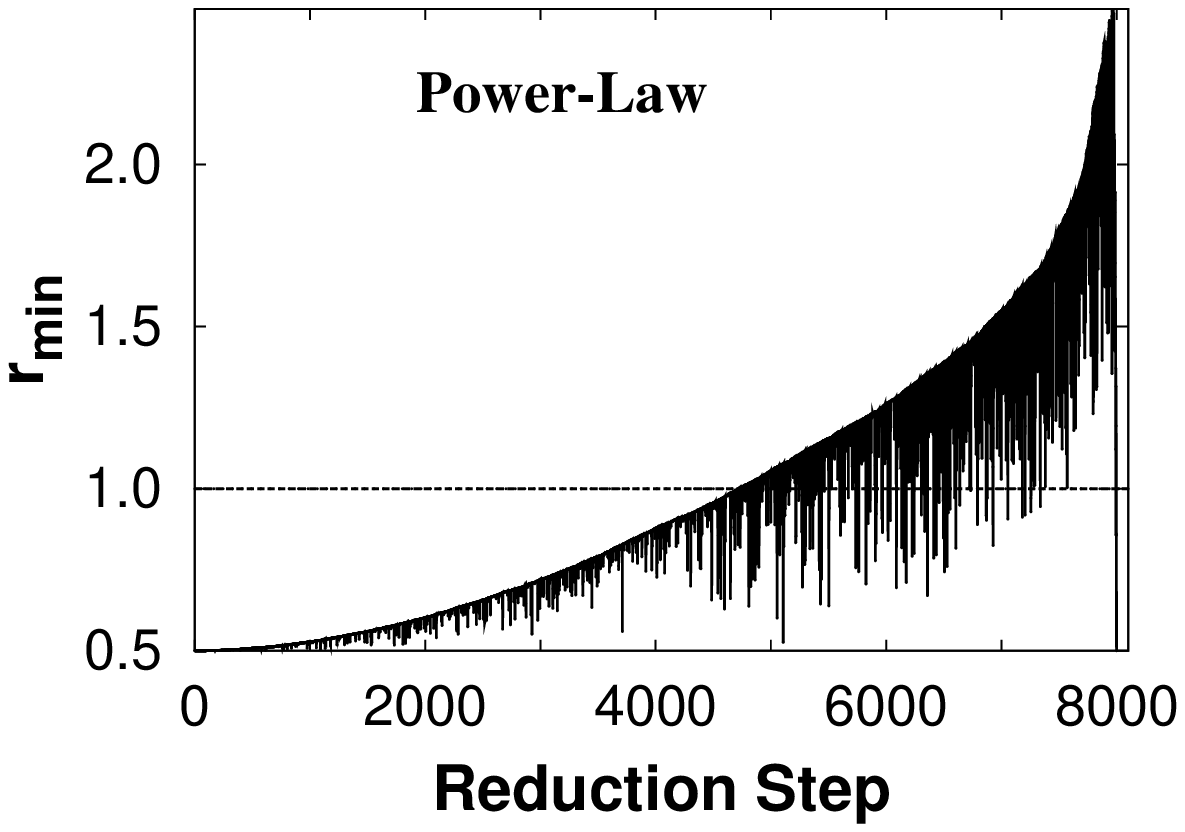}
\caption{\small Plot of $r_{min}$ during a run of DBR on a $N=20^3$
  lattice with Gaussian (top) and power-law bonds at $\gamma=1.5$ (bottom). }
\label{rplot}
\end{figure}

Like {\it Rule~VI\/} itself, DBR is not useful for homogeneous bond
distributions like $\pm J$, where all bonds have the same absolute
weight $|J|\equiv1$ (at least initially). But it may be very effective
for continuous bond distributions on dilute lattices with low average
connectivity $\langle\alpha\rangle$, as we have shown in
Sec.~\ref{numerics}. We can further exploit the universality of bond
distributions and utilize {\it broadly distributed} bonds~\cite{Cizeau93,Janzen08}, such as a
power law, $P(J)\sim |J|^{-\gamma}$ for
$|J|\to\infty$, as long as $P(J)$ has zero mean and finite width. In
fact, such a greedy procedure has already been described for extremely
widely separated bonds (each bond is larger in weight that the sum of
\emph{all} smaller ones), where it becomes
exact~\cite{Newman94,Cieplak94}: the problem is no longer NP-hard. For
the power-law bonds, we expect that there is a transition in the
behavior of this procedure at some finite value of $\gamma_c$. Hence,
we may find a ``sweet spot'': a sufficiently broad distribution on a
sufficiently dilute lattice for efficient DBR on very large lattices,
while $p$ just above $p_c$ and $\gamma$ just above $\gamma_c$ ensure
the EA universality class in any dimension.

Frustration leaves many bonds violated in the ground state, obviously,
and our recursive elimination procedure accounts for that through
compounding bonds in {\it Rule~III}, as described also in
Fig.~\ref{superbond}. In a simple benchmark, shown in
Fig.~\ref{2dextra}, we found that DBR obtains an approximate ground state energy
density of $\langle e\rangle_N=-1.308(1)$ for the undiluted EA, in
$d=2$ at $N=100^2$, only 0.5\% above $\langle
e\rangle_{\infty}=-1.31479(2)$, the best-known
result~\cite{Campbell04}, and $\langle e\rangle_N=-1.631(1)$ in $d=3$
at $N=20^3$, 4\% above $\langle
e\rangle_{\infty}=-1.700(1)$~\cite{Pal96b}. Computational costs are
trivial, $O(d\,N\ln N)$, but our implementation is limited to $N<10^4$
by a data structure problem: repeated application of {\it Rule VI}
leads to a few highly connected spins with hundreds of neighbors. Of
course, to calculate properties of low-$T$ excitations,
even a systematic
error of 4\% would be unacceptably large, and our naive DBR algorithm will have
to be developed into a meta-heuristic, again combining reduction and
EO.

To explore the effect of broadly distributed bonds, we have compared
DBR for one undiluted cubic lattice of size $N=20^3$ with Gaussian and
power-law bonds at $\gamma=1.5$. In Fig.~\ref{rplot}, we show the
value of the (smallest) $r_i=r_{min}$ in Eq.~(\ref{req}) of the
currently reduced spin $x_i$ during one run of DBR. Initially, for all
$r_{min}<1$, DBR is exact, which persists much longer for power-law
bonds. Even when $r_{min}\geq1$, the size of $r_{min}$ is typically
smaller for power-law bonds. Considering that DBR's systematic error
is only 4\% for Gaussian bonds, we project that power-law bonds should
be even more successful. We expect to conduct more extensive tests for
varying $\gamma$ and increasing $N$, which will require a
significantly revised data structure compared to the one used in these
studies.

\section{Conclusions}
\label{conclusions}
Our results validate the effectiveness of the recently proposed
reduction scheme to determine the stiffness exponent
$y$~\cite{MKpaper}, leaving remainder graphs that are a small fraction
(typically $\approx1-10$\%) of the size of the original problem in the
interesting regime just above $p_c$.  Note that the fact that
reduction works well in two-dimensions, where $T_c=0$ holds, does not
imply definitively that it should work for $d>2$. Nevertheless, since
the local interconnections between spins (i~e., graph vertices) near
and just above $p_c$ in $d=2$ and higher dimensions is similar,
justifies that the reduction scheme is applicable also for higher
dimensions~\cite{Boettcher04c,Boettcher04b,Boettcher05d}, or even
sparse random graphs~\cite{Boettcher03a,Boettcher03b}, where no exact
ground-state algorithms are available. As the general discussion and
the speculative inferences in the Appendix suggest, it may be possible
to extend the methods discussed here for any particular graph topology
or bond distribution at hand. We have also presented evidence that a
heuristic approach, based purely on the bond reductions introduced in
Sec.~\ref{reduction}, provides a fast algorithm to obtain approximate
ground states, with the potential to handle even large or undiluted
systems within bounded error.

\section*{Acknowledgments}
SB thanks P. Duxbury for inspiring discussions. This work was
supported by grant 0312510 from the Division of Materials Research at
the National Science Foundation and a grant from the Emory University
Research Council.

\section*{Appendix}
\label{appendix}
In general, a spin glass Hamiltonian $H$, such as the one in
Eq.~(\ref{Heq}), consists of  the sum (in the negative) of a number of terms
$J_{i_1,\ldots,i_p}x_{i_1}\ldots x_{i_p}$, each representing a
(hyper-)bond of weight $J$ between $p$ spins $x_i\in\{-1,+1\}$. The
number of connected spins $p$ may vary between terms, although $p=2$
for all terms in Eq.~(\ref{Heq}). The bonds $J$ are quenched variables
drawn from an arbitrary distribution, discrete or continuous.  A
particular {\it instance} of the spin glass Hamiltonian is specified
by the values of these quenched variables.

A ground state minimizes $H$, thus we want to maximize as many terms
as possible in $-H$.  Each spin, say $x_0$, appears in a number of
such terms, connecting it to a total of $q$ other spins. In general,
there are $2^q$ possible terms, and we can {\it eliminate} $x_0=\pm1$
by
\begin{eqnarray}
&&J_0x_0+J_1x_1x_0+\ldots+J_qx_qx_0+J_{12}x_1x_2x_0+\nonumber\\
&&\qquad\ldots+J_{1\ldots q}x_1\ldots x_qx_0\nonumber\\
&=&x_0(J_0+J_1x_1+\ldots+J_qx_q+J_{12}x_1x_2+\nonumber\\
&&\qquad\ldots+J_{1\ldots q}x_1\ldots x_q)\label{elimeq}\\
&\leq&|J_0+J_1x_1+\ldots+J_qx_q+J_{12}x_1x_2+\nonumber\\
&&\qquad\ldots+J_{1\ldots q}x_1\ldots x_q|\nonumber\\
&=&a_0+a_1x_1+\ldots+a_qx_q+a_{12}x_1x_2+\nonumber\\
&&\qquad\ldots+a_{1\ldots q}x_1\ldots x_q,\nonumber
\end{eqnarray}
where the bound again becomes an equality for the ground state energy.
A new Hamiltonian is obtained which is reduced by one variable. Notice
that the last two lines provide a unique system of $2^q$ linear
equations, one for each assignment of the $x_i=\pm1$, that determine
the new bonds $a$ in terms of the old bonds $J$.

To solve the linear system, we define
\begin{eqnarray}
g({\bf x})&=&|J_0+J_1x_1+\ldots+J_qx_q+J_{12}x_1x_2+\nonumber\\
&&\qquad\ldots+J_{1\ldots
q}x_1\ldots x_q|,
\label{gdefeq}
\end{eqnarray}
and note that for {\it any} function $g({\bf x}),~{\bf
x}\in\{\pm1\}^q$, it is true that
\begin{eqnarray}
g({\bf x})&=&\sum_{\{{\bf\epsilon}\}}
g({\bf\epsilon})\prod_{i=1}^q \delta_{x_i,\epsilon_i}\nonumber\\
&=&\sum_{\{{\bf\epsilon}\}}
g({\bf\epsilon})\prod_{i=1}^q\frac{1+\epsilon_ix_i}{2}
\label{gtransfeq}\\
&=&\sum_{\{{\bf\epsilon}\}}
\frac{g({\bf\epsilon})}{2^q}(1+\epsilon_1x_1+\ldots+\epsilon_qx_q+\epsilon_1\epsilon_2x_1x_2+\nonumber\\
&&\qquad\ldots+\epsilon_{q-1}\epsilon_qx_{q-1}x_q+\ldots+\epsilon_1...\epsilon_qx_1... x_q),\nonumber
\end{eqnarray}
where the sum extends over all $2^q$ permutations of
${\bf\epsilon}\in\{\pm1\}^q$ and the relation $\delta_{a,b}=(1+ab)/2$
for $a,b\in\{\pm1\}$ was used to represent the Kronecker
symbol. Comparison of Eq.~(\ref{gtransfeq}) with the last two lines in
Eq.~(\ref{elimeq}) yields
\begin{eqnarray}
a_{i_1,\ldots,i_p}=\frac{1}{2^q}\sum_{\{{\bf\epsilon}\}} g({\bf\epsilon})
\epsilon_{i_1}\ldots\epsilon_{i_p}
\label{newbondseq}
\end{eqnarray}
for the new bonds connecting the remaining variables.

In general, it is not useful to reduce the Hamiltonian in this way;
after all, if all $n$ spins are connected to each other, as for the
Sherrington-Kirkpatrick model~\cite{Sherrington75}, just the
elimination of {\it one} spin-variable requires $O(2^n)$
operations. Yet, there are common situations, in particular for
lattices of finite dimensionality and sparse graphs, where the
application of the previous results can be very useful: either in
itself, in combination with heuristic techniques, or as the basis of
approximation schemes. Of course, we may also be interested in the
system's entropy, the magnetization, overlaps, etc, which can be be
considered simultaneously~\cite{MKpaper}.

Although the combinatorial effort in the preceding expressions seems
daunting in general, they possess a rich structure that relate them to
other, well-studied subjects. For instance, we can rewrite
Eq.~(\ref{newbondseq}) as
\begin{eqnarray}
a_{i_1,\ldots,i_p}&=&\frac{1}{2^q}\sum_{j=0}^{2^q-1} g(\{j\})
  W_{k,2^q}(j)
\label{waveleteq1}
\end{eqnarray}
with
\begin{eqnarray}
k&=&1+\sum_{\mu=1}^p 2^{i_{\mu}-1},
\label{waveleteq2}
\end{eqnarray}
where $\{j\}$ is the binary encoding (on $\pm1$) of the integer
$j$. In particular, $W_{k,L}(x)$ is the $k$th Walsh function~\cite{Thuillard01} of
support $L$ familiar from wavelet analysis and
signal filtering. The orthogonality properties of Walsh functions
provide a powerful means to analyze the preceding reduction equations
for particular choices of initial bond distributions. For instance,
there may be types of graphs with a nontrivial bond distribution for
which the reductions could be simple. Also, a transformation may be
found that maps the distribution of the $J$'s into that of the
$a$'s. Finally, existing, highly optimized wavelet algorithms
\cite{Beylkin91} may in fact produce efficient spin glass solvers
based on these reduction equations.

\bibliographystyle{apsrev}
\bibliography{/Users/stb/Boettcher}

\end{document}